\documentclass[%
 reprint,
superscriptaddress,
 amsmath,amssymb,
 aps, prx,
]{revtex4-2}

\usepackage{graphicx}% Include figure files
\usepackage{dcolumn}% Align table columns on decimal point
\usepackage{bm}% bold math
\usepackage[colorlinks=true, linkcolor=red, urlcolor=blue, citecolor=red]{hyperref}
\usepackage{braket}
\usepackage{stackengine}
\usepackage{enumitem}
\usepackage{subfigure}

\begin{document}

\pagecolor{white}

\preprint{APS/123-QED}

\title{A Tensor Network Framework for Interpretable Graph Analysis of Brain Networks}

\author{Domenico Pomarico}
\affiliation{Dipartimento di Fisica, Università degli Studi di Bari Aldo Moro, I-70126 Bari, Italy}
\affiliation{Istituto Nazionale di Fisica Nucleare, Sezione di Bari, I-70125 Bari, Italy}

\author{Giuseppe Magnifico}
\email{giuseppe.magnifico@uniba.it}
\affiliation{Dipartimento di Fisica, Università degli Studi di Bari Aldo Moro, I-70126 Bari, Italy}
\affiliation{Istituto Nazionale di Fisica Nucleare, Sezione di Bari, I-70125 Bari, Italy}

\author{Alessandro Grecucci}
\affiliation{Dipartimento di Scienze della Formazione, Psicologia, Comunicazione (For.Psi.Com.), Università degli Studi di Bari Aldo Moro, I-70122 Bari, Italy}

\author{Loredana Bellantuono}
\affiliation{Istituto Nazionale di Fisica Nucleare, Sezione di Bari, I-70125 Bari, Italy}
\affiliation{Dipartimento di Biomedicina Traslazionale e Neuroscienze (DiBraiN), Università degli Studi di Bari Aldo Moro, I-70124 Bari, Italy}

\author{Jesus M. Cortes}
\affiliation{Biocruces-Bizkaia Health Research Institute, 48903 Barakaldo, Spain}
\affiliation{Biomedical Research Doctorate Program, University of the Basque Country, 48940 Leioa, Spain}
\affiliation{Department of Cell Biology and Histology, University of the Basque Country, 48940 Leioa, Spain} \affiliation{IKERBASQUE Basque Foundation for Science, 48009 Bilbao, Spain}

\author{Marianna La Rocca}
\affiliation{Dipartimento di Fisica, Università degli Studi di Bari Aldo Moro, I-70126 Bari, Italy}
\affiliation{Istituto Nazionale di Fisica Nucleare, Sezione di Bari, I-70125 Bari, Italy}

\author{Alfonso Monaco}
\affiliation{Dipartimento di Fisica, Università degli Studi di Bari Aldo Moro, I-70126 Bari, Italy}
\affiliation{Istituto Nazionale di Fisica Nucleare, Sezione di Bari, I-70125 Bari, Italy}

\author{Marlis Ontivero-Ortega}
\affiliation{Dipartimento di Fisica, Università degli Studi di Bari Aldo Moro, I-70126 Bari, Italy}
\affiliation{Istituto Nazionale di Fisica Nucleare, Sezione di Bari, I-70125 Bari, Italy}

\author{Alessandro Scarano}
\affiliation{Department of Psychology and Cognitive Science, University of Trento, Trento, Italy}

\author{Massimo Stella}
\affiliation{Department of Psychology and Cognitive Science, University of Trento, Trento, Italy}

\author{Roberto Bellotti}
\affiliation{Dipartimento di Fisica, Università degli Studi di Bari Aldo Moro, I-70126 Bari, Italy}
\affiliation{Istituto Nazionale di Fisica Nucleare, Sezione di Bari, I-70125 Bari, Italy}

\author{Sebastiano Stramaglia}
\affiliation{Dipartimento di Fisica, Università degli Studi di Bari Aldo Moro, I-70126 Bari, Italy}
\affiliation{Istituto Nazionale di Fisica Nucleare, Sezione di Bari, I-70125 Bari, Italy}

\author{Nicola Amoroso}
\affiliation{Istituto Nazionale di Fisica Nucleare, Sezione di Bari, I-70125 Bari, Italy}
\affiliation{Dipartimento di Farmacia - Scienze del Farmaco, Università degli Studi di Bari Aldo Moro, I-70125 Bari, Italy}

\date{\today}

\begin{abstract}

Identifying robust neurobiological signatures of brain disorders requires machine learning approaches that combine predictive performance with interpretable representations of feature interactions. Here we introduce a quantum-inspired framework based on tensor network machine learning that learns distributed representations of  gray-matter features encoded in a Matrix Product State representation, a variational ansatz originally developed for quantum many-body systems. The trained model is then used not only as a classifier but to extract quantum connected correlations between features, which encode higher-order feature interactions and which we map onto a weighted graph. This construction allows us to track, within a single representation, both: (i) the global spectral properties of the network (capturing collective learning dynamics), and (ii) node-level centrality measures (providing interpretable signatures of individual brain regions).
Using repeated train-test sampling schemes, we analyze two classification tasks on structural MRI data as examples of complex brain disorders: healthy controls versus schizophrenia and versus bipolar disorder. Node-level analysis identifies a stable set of gray-matter features, most prominently Heschl gyrus, insular cortex, and frontal regions, that act as hubs across multiple centrality measures and across resamplings. These centralities display lower variability across resamplings than Shapley values, supporting the interpretive value of the network representation. The bipolar feature set emerges as a subset of the schizophrenia one, consistent with the hierarchically organized neuroanatomical alterations reported in neuroimaging studies and offering a network-based characterization of this hierarchy.

\end{abstract}

\maketitle

%\tableofcontents

%\section{\label{sec:level1}First-level heading:\protect\\ The line
%break was forced \lowercase{via} \textbackslash\textbackslash}

\section{\label{sec1:level1}Introduction}

Recent advances in quantum computing have opened a stimulating frontier at the intersection of quantum technologies and machine learning \cite{arrazola, banchiscience, chinaphotonics, vakili2024, benedetti, hibat, Gili_2023, qgeneralize, holmes2024, Peters2023generalization,bowles2023contextuality, eisert2024}. Although current quantum processors remain limited by noise, decoherence, and restricted circuit depth, the conceptual framework of quantum information continues to inspire powerful computational models that can be implemented efficiently on classical hardware \cite{ibm_dqpt, entropy2, rydberg, PRXQuantum.2.010324, MPScircuit, Rudolph_2024, Miller_2024, schuhmacher2024hybridtreetensornetworks, Khosrojerdi_2025}. Among these, tensor networks, originally devised to study the properties of quantum many-body systems, such as quantum correlations and entanglement, have emerged as a versatile class of quantum‑inspired machine learning tools. Their ability to encode structured, higher-order correlations within a single global model, combined with their interpretable parameterizations, has positioned them as a complementary approach to conventional deep neural networks \cite{stoudenmire2017supervisedlearningquantuminspiredtensor, efthymiou2019tensornetworkmachinelearning, Huggins_2019, whaley, Felser_2021, Dborin_2022, Ballarin2023entanglemententropy, Collura2021, PhysRevX.8.011006, glasser2020, PhysRevB.101.075135, Gallego2022, PhysRevB.99.155131, chen2023machinelearningtreetensor, PhysRevResearch.4.043007}. Quantum-inspired approaches based on tensor networks have shown encouraging results in complex tasks such as cancer classification, biomarker discovery, and integrative analysis of gene‑expression networks \cite{Pomarico2025, repetto2024quantumenhancedstratificationbreast, breastQML}. 
These properties make tensor network models particularly attractive for neuroscience, where data are high-dimensional, noisy, and characterized by complex dependencies not easily captured by conventional statistical models~\cite{banchiscience, chinaphotonics, Omar2023, Magano2023, pharmaQML, recurrence, survival}. 

A central pillar of contemporary computational neuroscience is the modeling of brain organization through correlation-based measures, which quantify how activity or morphology co‑varies across regions \cite{AMOROSO201812, amoroso_age, la2023functional, Rizzi2025}. These approaches stem from the principle that the brain operates as a distributed system in which functional and structural components express coordinated fluctuations rather than acting in isolation \cite{Jimenez-Marin2024}. In functional Magnetic Resonance Imaging (fMRI), correlations between blood-oxygenation-level-dependent (BOLD) time series form the basis of functional connectivity estimates, allowing the identification of coherent networks that support perception, cognition, and affective regulation \cite{LOMBARDI2019150}. Similarly, in structural neuroimaging, covariance patterns, derived from correlations in gray‑ or white‑matter morphology across individuals, are interpreted as proxies of shared developmental, genetic, or experience‑dependent factors shaping brain architecture \cite{mri_data, YI2026112113}. This correlation‑based perspective  has proven particularly influential because it reveals network‑level dependencies that cannot be captured by local voxelwise comparisons alone, providing a more holistic depiction of interregional coordination underlying both healthy and pathological brain conditions \cite{amoroso_xai,Diez2025}. However, most covariance-based approaches remain fundamentally pairwise, potentially overlooking higher-order dependencies distributed across multiple interacting regions.

Beyond mapping intrinsic connectivity, correlation based models have become widely used tools for characterizing disease‑related alterations in network architecture. Changes in covariance patterns may reflect deviations in synchronized development, neurodegenerative processes, or dysregulation of large‑scale circuits implicated in psychiatric and neurological conditions \cite{JALALI2026120528, YI2026112113}. These network‑level anomalies may be more sensitive and interpretable than absolute volumetric changes, as they capture the reorganization of relationships among regions rather than their isolated deterioration \cite{YI2026121035}. As such, correlation‑driven modeling has become a key conceptual bridge between traditional neuroanatomy and computational network science, informing hypotheses about disconnectivity, compensatory pathways, and circuit‑level vulnerability. When combined with machine‑learning approaches, and especially with models capable of representing long‑range dependencies, such as tensor networks, correlation‑based measures can be used as structured inputs through which learning algorithms can infer latent organization principles, identify disease‑specific signatures, and generate hypotheses about large-scale organizational principles underlying brain dysfunction.

\begin{figure*}[t]
    \centering
    \includegraphics[width=1.00\linewidth]{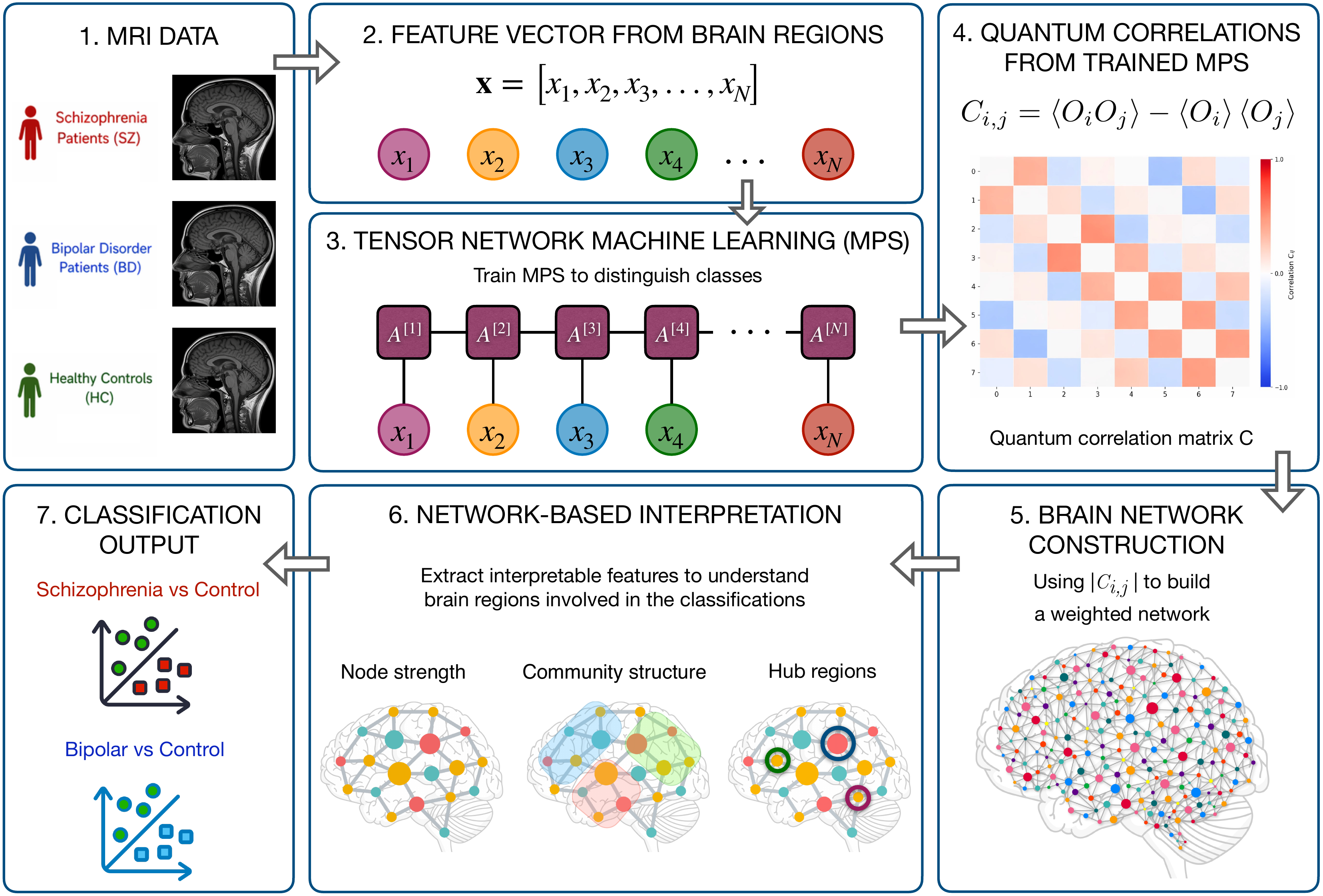}
    \caption{Schematic overview of the quantum-inspired tensor network framework for brain disorder classifications. (1) MRI data are considered for three groups: schizophrenia patients, bipolar disorder patients, and healthy controls. (2) Gray-matter concentration values extracted according to the AAL Atlas define a feature vector $\bf{x}$ for each subject. (3) Features are encoded into an MPS classifier, which is trained to distinguish diagnostic classes. (4) Quantum connected correlations are extracted from the trained MPS, yielding a pairwise correlation matrix. (5) The absolute values of the correlations are used as edge weights to construct a weighted network whose nodes are the gray-matter regions. (6) Graph-theoretic quantities are computed from the network to provide interpretable, region-level signatures of the learned representations. (7) The framework produces binary classification outputs for two diagnostic tasks: schizophrenia versus healthy controls and bipolar disorder versus healthy controls.}
    \label{fig:visual_scheme}
\end{figure*}

In this work, we combine these two perspectives. We train a tensor network, specifically a Matrix Product State (MPS) classifier, on structural MRI data, using it as both a predictive model and a probe of feature-level correlation structure. More in detail, we extract the quantum connected correlations of the trained MPS to define a weighted graph whose global spectral properties and node-level centralities provide, respectively, a dynamical and an interpretable view of the learning process. We apply this framework to two binary classification tasks: healthy controls versus schizophrenia, and healthy controls versus bipolar disorder. We find that a stable set of gray-matter features, most prominently Heschl gyrus, insular cortex, and frontal regions, emerges consistently across resamplings. Node-level centralities show substantially lower variability across resamplings than Shapley values, indicating that the network representation captures structural features of the data more robustly than model-based explanations. The set of discriminative features associated with bipolar disorder is found to be a subset of those features identified in schizophrenia, consistent with the hierarchically organized neuroanatomical alterations reported in neuroimaging studies, and in particular with previous structural covariance analyses on the same dataset showing broader network alterations in schizophrenia than in bipolar disorder \cite{mri_data}. The overall pipeline of the work is summarized in Fig.~\ref{fig:visual_scheme}.

The article is organized as follows. In Sec.~\ref{sec2:level1}, we introduce the structural MRI dataset and the tensor network learning framework used to encode gray‑matter features as quantum‑inspired representations, including our classification performance compared with a classical benchmark. In Sec.~\ref{sec3:level1}, we present the construction of the quantum pairwise‑correlation graphs that form the basis for our graph‑theoretic analysis. In particular, we detail how global spectral metrics and node‑level centrality measures extract interpretable neural signatures from the correlation structures, revealing stable hubs and task‑dependent reorganization patterns associated with the classification of schizophrenia and bipolar disorder relative to controls. In Sec.~\ref{sec4:level1}, we discuss the neuroscientific implications of these findings, emphasizing how quantum-inspired machine learning combined with correlation‑based graph modeling provides a principled, interpretable route for uncovering robust structural markers of brain disorders. Finally, in Sec.~\ref{sec:conclusion}, we summarize our findings.

\section{\label{sec2:level1}Materials and Methods}

\subsection{\label{secA:level2}Structural MRI dataset}

The dataset comprised 383 right‑handed adults (ages $18 - 65$), split into three demographically matched groups: schizophrenia ($n = 128$), bipolar I disorder ($n = 128$), and healthy controls ($n = 127$). Patients were recruited at Spanish clinical centers (Benito Menni CASM; Mare de Déu de la Mercè; Hospital Clínic de Barcelona). Controls were drawn from hospital non‑medical staff and the community and had no history of psychiatric illness or psychotropic treatment. Matching across groups was based on age, sex, and premorbid IQ (estimated via the Word Accentuation Test). Clinical characterization included PANSS for schizophrenia, and YMRS and HDRS for bipolar disorder; at scan time bipolar disorder participants included euthymic ($n = 77$), manic ($n = 28$), and depressed ($n = 23$) individuals. Common exclusion criteria were neurological disease, prior brain trauma, and alcohol/substance abuse within 12 months of imaging. All participants provided written informed consent under approval of the Comité de Ética de Investigación Clínica de las Hermanas Hospitalarias \cite{mri_data}.

High‑resolution structural MRI was acquired on a $1.5 T$ GE Signa system (General Electric, Milwaukee, WI). The T1‑weighted protocol used 180 axial slices, $1 \ mm$ slice thickness (no gap), $512 \times 512$ matrix, and $0.5 \times 0.5 \times 1 \ mm^3$ voxel size, with $TE = 4 \ ms$, $TR = 2000 \ ms$, and flip angle $15^\circ$.

Structural images underwent a standard pipeline: unified tissue segmentation in SPM12 to derive gray‑matter partial‑volume maps; brain extraction per BET; and spatial normalization to the MNI152 ($2 \ mm$) template using FSL registration tools. Deformation fields from normalization were then applied to generate normalized gray-matter images, which were down‑sampled to $4 \times 4 \times 4 \ mm^3$ to reduce computational burden while preserving the large-scale spatial covariance structure relevant for network-level analyses. Grey matter concentration for each region was extracted according to the AAL Atlas. This regional representation was used to characterize distributed structural variation across subjects rather than localized voxel-wise abnormalities.

\subsection{\label{secB:level2}Tensor network machine learning}

We employ Matrix Product States (MPS), a tensor network ansatz that provides a compact and controllable parameterization of high-dimensional model tensors while preserving sufficient expressive power for supervised classification tasks. An MPS represents a tensor with many indices as a contracted sequence of three-index tensors arranged along a one-dimensional chain. The bond dimension~$\chi$ of the MPS controls the maximum allowed quantum correlations and entanglement across the chain and therefore the representational capacity \cite{PhysRevLett.69.2863, RevModPhys.77.259, SCHOLLWOCK201196, RevModPhys.93.045003}.

Given an input sample with $N$ features, we embed the data into a product state of $N$ qubits arranged on a one-dimensional lattice. Each qubit encodes a single feature through a fixed local embedding $\ket{\phi(x^{(j)})}=\left( \cos(x^{(j)}), \ \sin(x^{(j)}) \right)^\intercal\in\mathbb{R}^2, j=1, \dots, N$, which is the standard sinusoidal feature map of MPS-based classification problems \cite{stoudenmire2017supervisedlearningquantuminspiredtensor}. For a binary classification task with class index $\ell\in\{0, 1\}$, the MPS defines the predictor as the contraction of the weight tensor $W^{\ell}$ with the product feature state
\begin{equation}
f_W^{\ell}(x)
=
\sum_{s_1,\ldots,s_N}
W^{\ell}_{s_1,\ldots,s_N}\;
\ket{\phi(x^{(1)})_{s_1}}
\otimes\cdots\otimes
\ket{\phi(x^{(N)})_{s_N}},
\label{eq:predictor}
\end{equation}
and is trained by minimizing the mean-squared-error cost
\begin{equation}
\mathcal{C}(W)
=
\frac{1}{2N_T}
\sum_{\omega=1}^{N_T}
\sum_{\ell}
\left(
f_W^{\ell}(x_\omega)
-
y^{\ell}_{\omega}
\right)^2,
\label{eq:mse}
\end{equation}
where $N_T$ is the number of training samples and $y^{\ell}_{\omega}$ denotes the one-hot encoded label. The predicted class corresponds to the index $\ell$ maximizing the magnitude of $f_W^{\ell}(x)$ \cite{stoudenmire2017supervisedlearningquantuminspiredtensor,chen2023machinelearningtreetensor}.

Training proceeds via gradient descent using a two-site update scheme: at each step, two adjacent MPS tensors are jointly optimized and then split via a truncated singular-value decomposition that enforces the bond-dimension constraint and retains the most informative components of the model~\cite{stoudenmire2017supervisedlearningquantuminspiredtensor, efthymiou2019tensornetworkmachinelearning, Huggins_2019, whaley, Felser_2021, Dborin_2022, Ballarin2023entanglemententropy, Collura2021, PhysRevX.8.011006, glasser2020, PhysRevB.101.075135, Gallego2022, PhysRevB.99.155131, chen2023machinelearningtreetensor, PhysRevResearch.4.043007,guo2021neuraltangentkernelmatrix}. The two-site optimization moves back and forth along the one-dimensional lattice; a full pass that returns the optimization window to its initial position defines a \emph{sweep}, conceptually analogous to a training epoch. The resulting training dynamics can exhibit a dynamical entanglement transition that shares qualitative features with the grokking phenomenon recently reported in tensor network classifiers~\cite{pomarico2025grokkingentanglementtransitiontensor,technologies13100438}, and that we analyze in Appendix \ref{app:sec1}.

To track the learning dynamics, we compute several quantum observables on the MPS: the reduced density matrix in label space $\varrho^{(\ell)}$, the local magnetizations $\langle\sigma^{k,i}_Z\rangle$ for feature $i$ and label subspace $k\in\{0,1\}$, and the pairwise correlation functions $\langle\sigma^{k,i}_Z \sigma^{k',j}_Z\rangle$. From these, we define the connected correlation measure
\begin{equation} C^{(k,k')}_{i,j}
=
\frac{
\langle \sigma^{k,i}_Z \sigma^{k',j}_Z \rangle
-
\langle \sigma^{k,i}_Z \rangle\,\langle \sigma^{k',j}_Z \rangle
}{
\sqrt{\varrho^{(\ell)}_{k,k}
\varrho^{(\ell)}_{k',k'}
}},
\qquad
k,k'\in\{0,1\},
\label{eq:connectedcorr}
\end{equation}
which leads to the identification of feature pairs that contribute most prominently to the classification process.

Finally, we characterize the entanglement structure along the lattice by evaluating the class-resolved von Neumann entropies across the generic bond $(i,i+1)$, that is the quantity $S^\ell(i)=-\textrm{Tr} ( \varrho^{(\ell)}_{1:i} \textrm{ln} \varrho^{(\ell)}_{1:i} )$, where $\varrho^{(\ell)}_{1:i}$ is the reduced density matrix of the MPS partition comprising sites $1,2,...,i$ \cite{RevModPhys.80.517, RevModPhys.82.277}. Monitoring these quantities across training reveals the transition from the initial volume-law entanglement of the random MPS towards a sub-volume scaling regime, accompanied by spectral reorganization in the reduced density matrices that reflects emerging feature selectivity \cite{pomarico2025grokkingentanglementtransitiontensor,technologies13100438}. We define the transition sweep as the sweep at which these internal observables, i.e. magnetization patterns, reduced density matrix coherence, entanglement entropy, undergo simultaneous sharp reorganization. %, marking the onset of stable generalization.

\subsection{Classification performance}\label{sec:classification_performance}

\begin{figure}[t]
\includegraphics[width=0.95\linewidth]{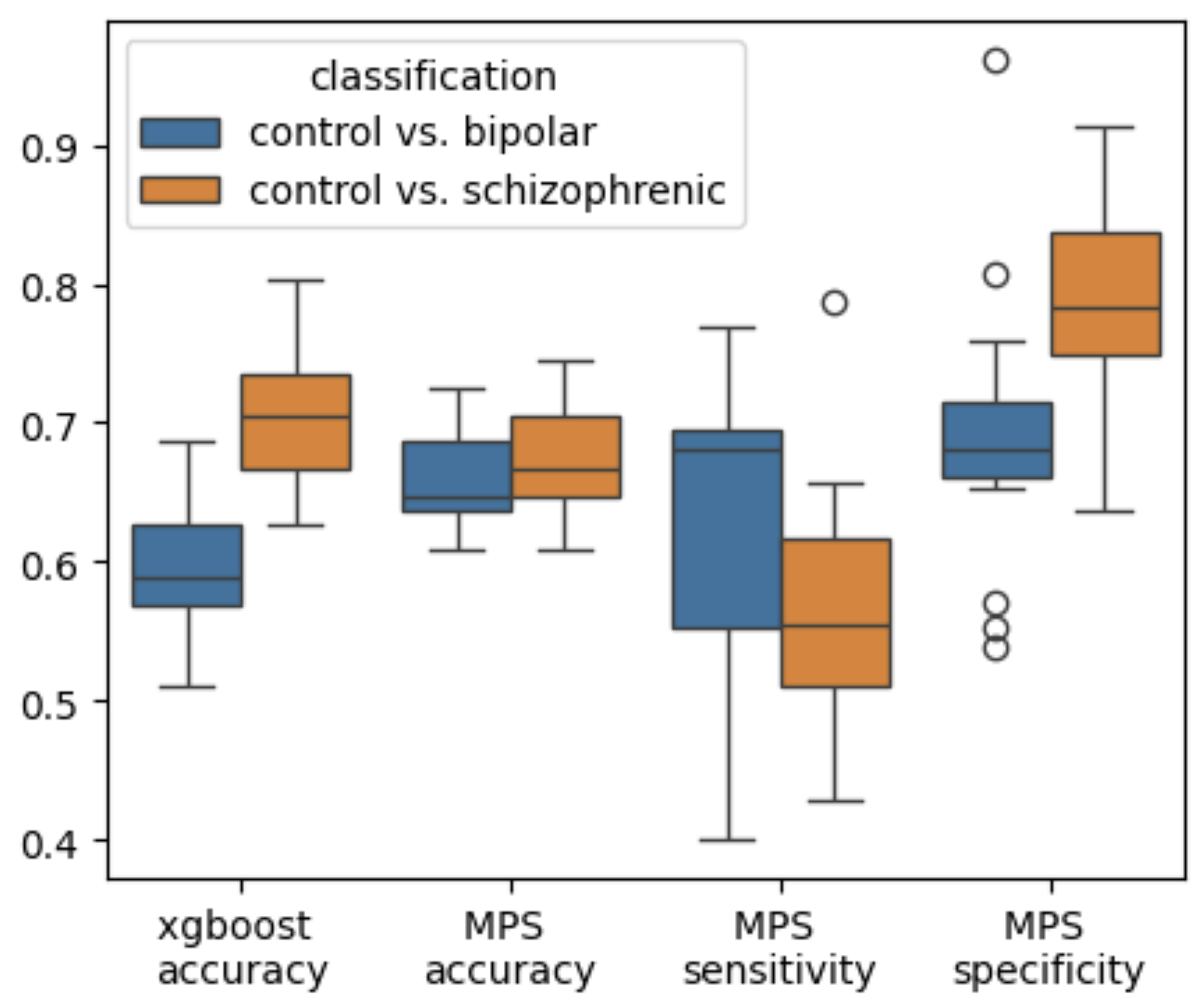}\caption{\label{fig:boxplot}Boxplot comparison of classifier performance across the two diagnostic tasks.
The distribution of performance metrics are obtained for an XGBoost classifier and the Matrix Product State (MPS) machine‑learning model when distinguishing control vs. bipolar subjects (blue) and control vs. schizophrenic subjects (orange). Shown are boxplots for XGBoost accuracy, MPS accuracy, MPS sensitivity, and MPS specificity.}
\end{figure}

We compared the MPS classifier with a tree‑ensemble benchmark (XGBoost), chosen as a solid, hyperparameter-robust baseline, on two binary learning tasks: controls vs. bipolar disorder and controls vs. schizophrenia. Both tasks were derived from $N=30$ gray‑matter features selected via Boruta ranking \cite{JSSv036i11}, and performance was assessed over 15 random train/test resamples (test proportion $0.20$). Fig. \ref{fig:boxplot} reports the resulting distributions of accuracy for both methods, together with sensitivity and specificity for MPS, evaluated at the best performing sweep during training. 
%\textcolor{red}{(GM: è il transition sweep?)} \textcolor{blue}{non necessariamente}

For XGBoost, the control vs. schizophrenia task yielded clearly higher accuracy than control vs. bipolar, with a median around $0.71$ (interquartile range IQR $\approx [0.67, 0.74]$) versus 0.59 (IQR $\approx [0.57, 0.63]$). MPS exhibited more balanced performance across tasks: for bipolar, the median accuracy was $0.65$ (IQR $\approx [0.64, 0.69]$), exceeding XGBoost by roughly six percentage points; for schizophrenia, MPS reached $0.67$ (IQR $\approx [0.65, 0.71]$), slightly below XGBoost. The bipolar classification task exhibits a statistically significant difference in accuracy distributions ($p \approx 0.001 < 0.01$), whereas the schizophrenia classification task does not show statistically significant discrimination ($p \approx 0.1 > 0.01$). Two patterns emerge: (i) the control vs. schizophrenia separation is intrinsically easier than control vs. bipolar for the given features, as both classifiers reach higher accuracy on the former; and (ii) MPS captures structure that benefits the bipolar task in particular, where it improves on the tree-ensemble baseline.

For MPS, sensitivity (recall of the patient class) was higher in the bipolar task (median $0.68$, IQR $\approx [0.55, 0.70]$) than in the schizophrenia task (median $0.56$, IQR $\approx [0.51, 0.62]$, with occasional higher outliers). Specificity displayed the opposite asymmetry, favoring the schizophrenia task (median $0.78$, IQR $\approx [0.75, 0.84]$, with high outliers around 0.90) over the bipolar one (median $0.68$, IQR $\approx [0.66, 0.71]$, with a few low outliers near 0.55). 

This pattern suggests that MPS tends to reduce false positives on the schizophrenia task (strong specificity) at the expense of higher false negatives (lower sensitivity), whereas performance on the bipolar task is more balanced between sensitivity and specificity. Considering the interquartile ranges, accuracy distributions for both models are comparatively tight, indicating stable central performance under data splits. Sensitivity displays greater dispersion, particularly for MPS, implying that fold composition materially affects recall of the patient class. Overall, MPS achieves accuracy that improves over the XGBoost baseline on the harder bipolar task and matches it on the schizophrenia task. 

A central motivation for using quantum-inspired tensor network architectures is their capability to yield interpretable representations of feature interactions. As discussed in Appendix~\ref{sec:quantum_inspired_interpretability}, the single-site entanglement entropy displays a broadly uniform pattern across features in both tasks, with consistently large values and limited variability across runs. Its uniformity indicates that the stable features are all significantly integrated into the collective quantum correlations of the classification structure, providing a quantum-inspired validation of the feature selection: the stable selected regions are not only statistically discriminative but also exhibit a high level of quantum correlations in the global representations. This stability supports the use of quantum-inspired observables as reliable measures of feature integration in the learned representation, and motivates the network-based analysis presented in the following sections, where we move from single-site quantum properties to pairwise quantum correlations between features, constructing a weighted brain network whose topology encodes the learned classification structure.

\section{\label{sec3:level1}Quantum pairwise correlation graph}

The connected correlation measure defined in Eq.~\eqref{eq:connectedcorr} assigns, to each pair of features $(i,j)$ and each pair of label indices $(k,k')$, a real-valued quantity $C^{(k,k')}_{i,j}$ computed from the trained MPS. We interpret the absolute value of these quantum correlations $ \left | C^{(k,k')}_{i,j} \right |$ as edge weights of a weighted graph whose nodes are the gray-matter features; these weights represent learned interaction strengths within the classifier rather than direct anatomical or functional connectivity. (Fig.~\ref{fig:visual_scheme}, steps 4–5). This defines three graphs of interest: two diagonal ones, $ \left | C^{(0,0)} \right |$ and $ \left | C^{(1,1)} \right |$, capturing intra-class correlations for the control and pathological subspaces respectively, which we refer to as the control mask and the pathology-associated mask; and the off-diagonal $\left | C^{(0,1)} \right | $, capturing cross-class correlations, which we refer to as the confounding mask. The three graphs evolve during training and encode complementary aspects of the learned representation.

In the following sections, we analyze the evolution of the three graphs during training through global spectral observables, which capture collective reorganization of the correlation structure, and through node-level centrality measures at the transition sweep, which identify the regions that act as hubs of the learned representation.

\subsection{Global spectral observables}

\begin{figure*}
\includegraphics[width=0.95\linewidth]{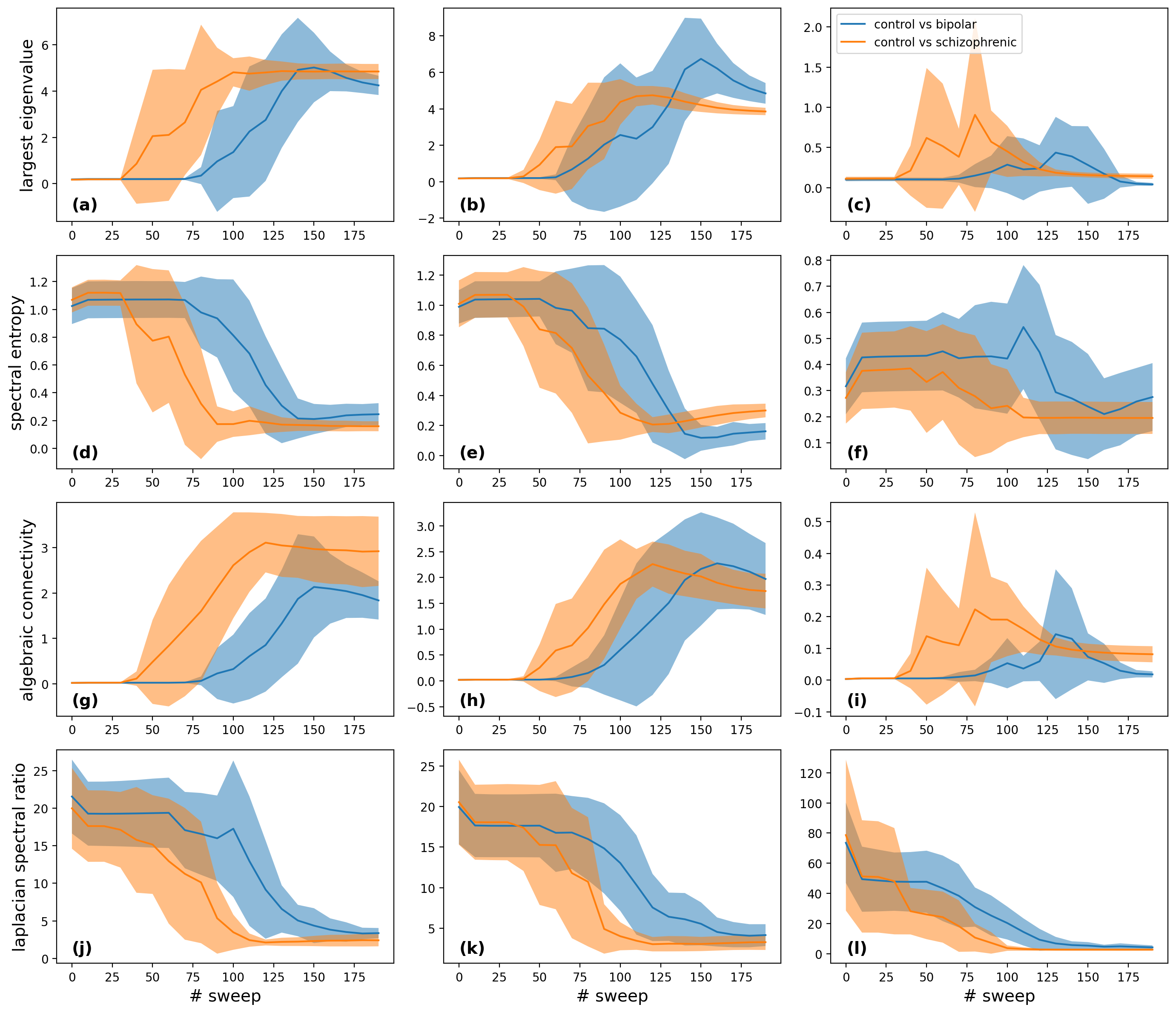}
\caption{\label{fig:global}Evolution of graph-theoretic and spectral metrics computed from the pairwise quantum correlations matrices $\left | C^{(k,k')}_{i,j}\right |$ during training. Each row reports a different network metric: largest eigenvalue (a–c), spectral entropy (d–f), algebraic connectivity (g–i), and Laplacian spectral ratio (j–l), tracked over training sweeps. Columns correspond to (left) the control mask $\left | C^{(0,0)} \right |$, (center) the pathological mask $\left | C^{(1,1)} \right |$, and (right) confounding mask $\left | C^{(0,1)} \right |$. For each panel, curves show the mean metric value across sweeps, and shaded regions indicate the standard deviation across the train-test resampling runs. The two binary classification tasks are shown in blue (control vs. bipolar) and orange (control vs. schizophrenia).}
\end{figure*}

Once we have established that the MPS classifier captures discriminative structure in both tasks, we can examine how this structure emerges during training in the quantum pairwise correlation network. We monitored a set of global graph-theoretic and spectral observables derived from the correlation matrices $C^{(0,0)}$, $C^{(1,1)}$, $C^{(0,1)}$. Further details on the definition of the global graph-theoretic and spectral observables are provided in Appendix~\ref{app:sec2}. 

The temporal evolution of these  observables across training sweeps reveals clear signatures of the dynamical transition, an abrupt reorganization of the internal correlation structure, coinciding with a specific transition sweep for each task, where the model starts to generalize properly. This transition is clearly visible in Fig.~\ref{fig:global}, which shows the four metrics we analyze: the largest eigenvalue of the adjacency matrix, the spectral entropy, the algebraic connectivity of the Laplacian, and the Laplacian spectral ratio as a function of the sweep number. Further details on this dynamics are discussed in Appendix~\ref{app:sec1}.

Together, these quantities probe complementary aspects of the network topology, including global coherence, spectral complexity, connectivity robustness, and synchronization capacity. We refer to global spectral quantities as topological invariants of the weighted graph in the sense that they are invariant under node relabeling and depend only on the underlying connectivity structure.

The largest eigenvalue of the adjacency matrix provides a measure of the global strength and coherence of correlations within the network \cite{Chung1997}. As shown in panels (a-c), this quantity exhibits a marked increase during training sweeps for both classification tasks, signaling the progressive emergence of a dominant correlation mode within the learned representation. Importantly, growth occurs earlier for the schizophrenia task than for the bipolar task. This suggests that both control and pathological masks associated with schizophrenia classification develop structured correlations earlier during training. In contrast, bipolar classification requires a longer training period before the network reorganizes, producing a delayed increase in the leading eigenvalue. After reaching a peak, the eigenvalue tends to stabilize, indicating that the system transitions from a phase dominated by growing correlations to a more stable configuration in which correlations are redistributed across multiple modes. Notably, the confounding mask case in panel (c) exhibits a distinct non-monotonic evolution of the largest eigenvalue, characterized by a transient increase followed by a decline, suggesting the formation of short-lived collective correlations that do not stabilize into a persistent global structure.

The spectral entropy of the adjacency spectrum quantifies the diversity of modes that contribute to the network structure \cite{Thomas_2025}, with lower values indicating dominance by a few modes. Panels (d–f) show a pronounced drop during training for both diagonal masks, a spectral condensation in which the correlation structure collapses onto a small number of dominant modes. As with the largest eigenvalue, the drop occurs earlier for the schizophrenia task, providing consistent evidence that the transition is delayed in the bipolar task. The confounding mask in panel (f) instead maintains relatively comparable and broadly distributed entropy values throughout the training, indicating that no comparable spectral condensation takes place.

The algebraic connectivity measures the global connectedness and robustness of the network \cite{Fiedler1973, DEABREU200753}. Larger values indicate stronger integration and greater resistance to partitioning into disconnected communities. Panels (g–i) reveal a rapid increase in algebraic connectivity during training for both diagonal masks. This suggests that the network transitions from a relatively fragmented correlation structure to a highly connected configuration. The timing of this increase again differs between the two classification problems: the rise occurs earlier for the schizophrenia task and later for the bipolar one, a temporal shift that provides a topological signature of the delayed organization. In contrast, the confounding masks display a non-monotonic evolution of the algebraic connectivity, with a transient increase followed by a reduction, indicating the temporary formation of integrated correlation pathways that subsequently weaken rather than stabilizing.

The Laplacian spectral ratio characterizes the spectral spread of the Laplacian and the dynamical stability of the network \cite{PhysRevLett.89.054101, PhysRevE.77.031102, YOU20121245}, with small ratios indicating improved synchronization capacity and structural regularity. Panels (j–l) show a dramatic reduction of this ratio as training progresses, indicating that the network becomes more homogeneous and dynamically stable. The transition again occurs earlier in the schizophrenia classification than in the bipolar one. The convergence of the ratio toward small values suggests that both tasks reach a regime of strong global coordination, but the onset of this regime differs across training sweeps. It is possible that this difference reflects the greater extent of cortical alterations in schizophrenia compared to bipolar disorder. In particular, Grecucci et al. \cite{mri_data}, using the same dataset, showed that schizophrenia was characterized by multiple significantly altered structural networks relative to controls, whereas bipolar disorder exhibited a much more limited pattern of network differences. This asymmetry in network-level disruption may contribute to the earlier emergence of structured correlations observed in the schizophrenia classification task, as observed at the level of effective tensor network representations.

These spectral observables provide a consistent and coordinated description of the training dynamics. Importantly, these dynamics should be interpreted as changes in the internal organization of the learned representation rather than as temporal changes in biological brain networks. Initially, the quantum correlation networks display relatively weak structure imposed by random initial conditions, characterized by moderate eigenvalues, high spectral entropy, low algebraic connectivity, and large Laplacian spectral ratios. As training progresses, the networks undergo a sharp topological reorganization marked by the dominant eigenmode growing, the spectral entropy collapsing, the algebraic connectivity rising, and the Laplacian spectrum compressing.

This coordinated set of changes identifies the transition in which the internal representations become globally structured and coherent. Crucially, the timing of this transition differs between classification tasks: the pathological masks associated with schizophrenia reach the organized regime earlier, while the bipolar case exhibits a systematic delay in spectral reorganization. The confounding cross cases display weaker and less structured transitions, indicating that the spectral signatures observed in the other panels are specifically associated with meaningful pathological discrimination.

By capturing large-scale reorganization of the learned correlation structure, these observables show how the MPS classifier, through its quantum correlations, progressively assembles the gray-matter features into a coherent network of brain regions whose topology encodes the discriminative signatures of the classification task.

\subsection{Node-level interpretable quantities}\label{sec:node_level_quantities}

\begin{figure*}
\includegraphics[width=0.7\linewidth]{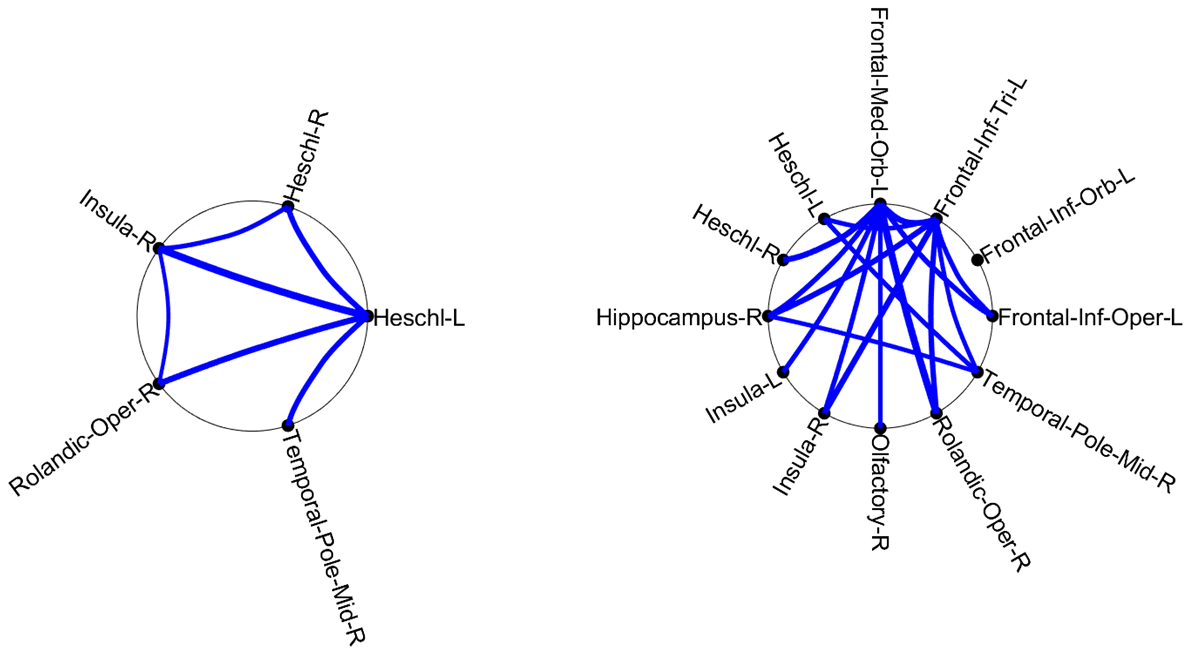}
\caption{\label{fig:bip_sz_cp} Circular plot of pairwise regions with significant differences in correlation values between controls and patients groups:  bipolar disorder (left), and schizophrenia (right).}
\end{figure*}

Once the global topological reorganization of the correlation network obtained from quantum observables has been characterized, we identified the individual brain regions that act as hubs of this reorganization, focusing 
on the same stable feature sets considered in Appendix~\ref{sec:quantum_inspired_interpretability}: 12 regions for schizophrenia and 5 for bipolar, i.e. the features consistently selected by Boruta across all 15 resampling runs. Importantly, the five features selected in the bipolar classification are a subset of the twelve features identified in the schizophrenia task, indicating that the latter classification involves a broader but overlapping network of discriminative regions, in line with previous findings \cite{mri_data}.

Figure \ref{fig:bip_sz_cp} shows the pairwise regions that exhibit statistically significant differences in correlation values between the control and patient groups after Bonferroni correction for multiple comparisons. Before the two-sample t-tests, the correlation networks from all training resampling were transformed into Fisher’s z-scores. Notably, all significant connections reflect increased correlation between regions in both patient groups, i.e. bipolar disorder and schizophrenia, compared to controls (see Fig. \ref{fig:bip_sz_mt} in Appendix \ref{app:sec3}). These increases should be interpreted within the learned interaction topology of the classifier rather than as direct evidence of increased biological connectivity. This pattern suggests that the quantum-inspired algorithm focuses on pathological feature correlations. The left Heschl gyrus for the bipolar disorder task, and the frontal medial orbital and the frontal inferior triangular gyrus for the schizophrenia task, are the regions with the highest number of significant connections indicating that these regions emerge as highly central nodes within the learned interaction structure associated with classification.

\begin{figure*}
\stackunder[5pt]{\includegraphics[width=0.8\linewidth]{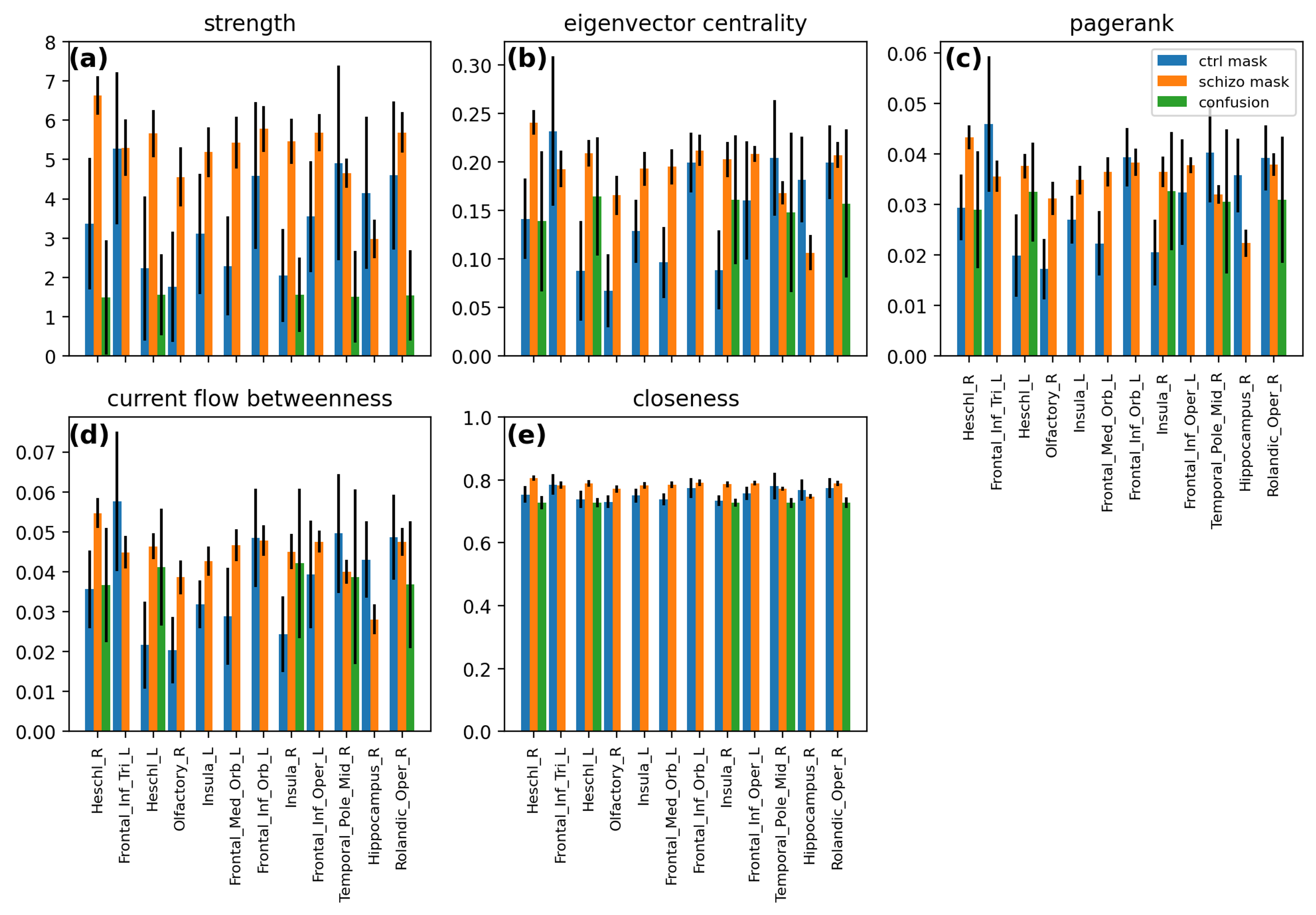}}{}
\stackunder[5pt]{\includegraphics[width=0.8\linewidth]{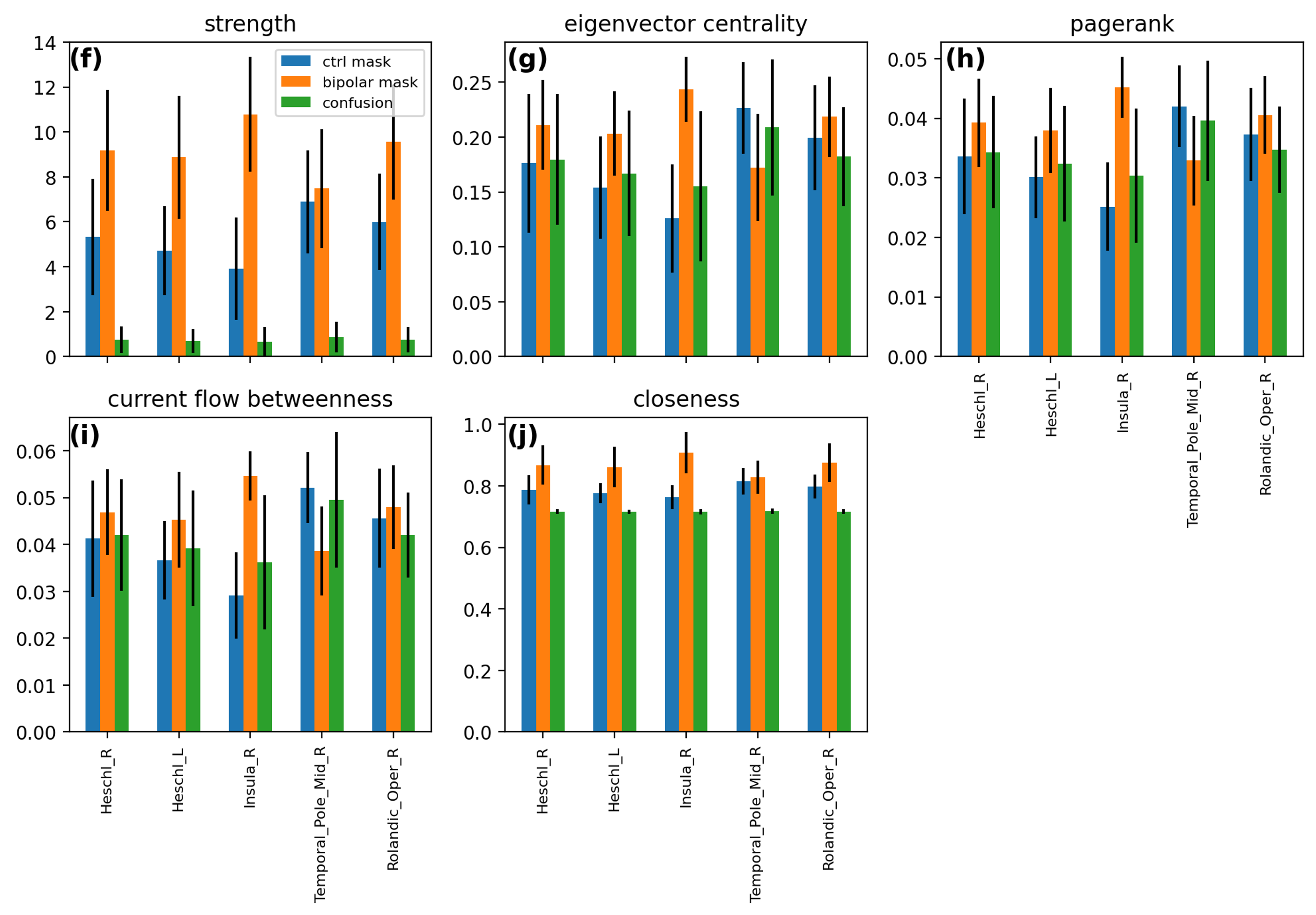}}{}
\caption{\label{fig:node1}Comparison of node‑level graph metrics derived from pairwise correlation matrices at the transition sweep for the control vs. schizophrenia classification in (a-e) and control vs. bipolar in (f-j). Panels report node‑level strength, eigenvector centrality, PageRank, current‑flow betweenness, and closeness computed from the control, pathological, and confounding pairwise correlation matrices (blue, orange, and green bars, respectively). The analysis focuses exclusively on the brain regions that were consistently selected across all 15 resampling runs following Boruta feature‑ranking, ensuring that only the most stable and high‑importance regions are retained.}
\end{figure*}

Now we investigate the node-level importance, by focusing on graph-theoretic centralities which provide a topological characterization of how individual brain regions contribute to the classification. Fig. \ref{fig:node1} reports five node-level centrality measures of the correlation networks derived from trained quantum states, i.e. strength, eigenvector centrality, PageRank, current-flow betweenness, and closeness, defined in Appendix \ref{app:sec2}, computed over the 15 resampling runs.

Panels (a–e) correspond to the classification controls vs. schizophrenic patients and focus on the 12 features shared across all runs. Panels (f–j) correspond to controls vs. bipolar patients and focus on the 5 features consistently shared across the same number of samplings.  For each feature, we compare the stability of its graph-theoretic centralities across runs with that of the Shapley attribution scores~\cite{shapley1951notes,lundberg2017unified} derived from the XGBoost tree-ensemble baseline introduced in Sec.~\ref{sec:classification_performance}. The Shapley values, shown in Fig.~\ref{fig:entropy_feature_importance}(b,d) in Appendix~\ref{sec:quantum_inspired_interpretability}, provide a complementary, classically computed view of feature importance (independent of the quantum correlations of the MPS).

Strength (panels a and f) quantifies the total connectivity of a node within the correlation network, i.e. how strongly a feature co-varies with the others. In the schizophrenia task, the pathological mask exhibits systematically larger strengths than the control mask for several brain regions, most prominently the Heschl gyrus, the frontal inferior opercular and triangular regions, and the insula; conversely, the hippocampus emerges as a relatively more stable integrative node in the control mask. In the bipolar task, the five shared features display the same pattern: elevated strength for the pathological mask relative to the control. The confounding mask shows markedly reduced strength across all regions, indicating weak and inconsistent cross-class correlations. The narrow error bars of the pathological masks across the 15 resamplings confirm the stability of these observations, which are consistent with previous findings on the same dataset \cite{mri_data} reporting more extensive structural covariance alterations in schizophrenia than in bipolar disorder.

Panels (b-d) and (g-i) report eigenvector centrality, PageRank, and current-flow betweenness. Despite their different formal definitions, capturing, respectively, embedding in densely connected neighborhoods, diffusion accessibility, and mediation of correlation flow, the three measures produce consistent rankings across nodes and masks. In the schizophrenia task, the pathological mask assigns the highest centrality to the frontal inferior triangular gyrus, the temporal pole, the Heschl gyrus, and the Rolandic operculum; the control mask additionally emphasizes the hippocampus. In the bipolar task, the shared features retain moderate-to-high centrality values, with an overall organization less polarized than in the schizophrenia case. The confounding mask displays lower and less structured centrality patterns throughout. The agreement among three measures with distinct analytical origins, together with the narrow error bars across the 15 resamplings, indicates a stable correlation backbone where the
identified regions jointly act as hubs, diffusion centers, and bridges of correlation flow within the network.

Closeness quantifies how efficiently information can spread from a node to the rest of the graph via shortest paths \cite{Newman2018}, as expressed by panels (e) and (j). The closeness values show limited variation between masks and across runs. This behavior indicates that, although the relative strengths of correlations change, the global geometric accessibility of these regions remains relatively stable within the correlation network. The highest values tend to occur for Heschl gyrus, frontal inferior triangular gyrus and rolandic operculum. This indicates that these regions occupy central geometric positions in the correlation network, and this property remains relatively stable regardless of the mask.

The network-based centrality measures display substantially lower variability across the 15 resampling runs than the Shapley values reported in Appendix~\ref{sec:quantum_inspired_interpretability} (Fig.~\ref{fig:entropy_feature_importance}), confirming that the topology emerging from the quantum correlations of the MPS provides a more robust characterization of the underlying relational structure among features.

\section{\label{sec4:level1}Discussion}

The node-level analysis of the quantum pairwise correlation networks offers a direct interpretable link between the topological organization of gray matter features and the neurobiological substrates underlying schizophrenia and bipolar disorder. Importantly, these networks should be interpreted as model-derived interaction structures reflecting distributed classification-relevant organization rather than direct anatomical connectivity. By focusing on features that remain stable across 15 independent training–test samplings, the analysis isolates regions whose importance does not depend on a specific realization of the dataset but instead reflects persistent structural relationships in the gray matter correlation network within the learned representation, which is consistent with the broader perspective that structure-function interplay is fundamental to understanding disease mechanisms \cite{10.1162/NETN.a.22}. 

This pattern aligns with previous findings on the same dataset, where these regions participated in distributed structural covariance networks, with a broader involvement in schizophrenia and a more circumscribed engagement in bipolar disorder \cite{mri_data}. Importantly, these regions should be interpreted as components of distributed networks rather than isolated loci, suggesting that the identified features reflect stable hubs within a shared structural backbone within the learned interaction topology.

A key observation is the nested structure of the discriminative feature sets. The schizophrenia classification task identifies a broader set of twelve stable regions, whereas the bipolar classification isolates a reduced subset of five regions, all of which are contained within the schizophrenia feature set. This hierarchical organization is consistent with previous findings on the same dataset \cite{mri_data}, where schizophrenia was associated with a more extensive pattern of structural covariance alterations across multiple networks, while bipolar disorder exhibited a more circumscribed configuration embedded within a partially overlapping set of regions. These results are also supported by large meta-analyses that reported that both disorders involve gray matter reductions in frontal and insular regions \cite{bipolar-schizo, frontal}, but schizophrenia typically displays more extensive cortical, sensorimotor and cognitive alterations compared with bipolar disorder \cite{SUZUKI200241, SORELLA2019101854}, pointing to a broader principle of heterogeneity in brain organization that extends to other neurodevelopmental disorders \cite{rasero2023}. 

Several of the most stable nodes identified in our analysis correspond to cortical regions repeatedly implicated in the pathophysiology of psychotic disorders: 
\begin{enumerate}[label=(\roman*)]
    \item The presence of Heschl gyrus among the most stable nodes is consistent with extensive evidence linking anomalies of the primary auditory cortex to schizophrenia. Structural MRI studies show significant reductions in gray matter volume in Heschl gyrus in patients with schizophrenia compared with both healthy controls and bipolar disorder patients \cite{heschl, heschl2}. These alterations are believed to contribute to the abnormal auditory processing and hallucination-related symptoms that characterize the disorder \cite{auditory}. The elevated network centrality of this region within the learned interaction network is consistent with the prominent involvement of auditory cortical systems reported in schizophrenia literature.
    \item Another recurrent region in the shared feature sets is the insula, which plays a key role in interoception, emotional regulation, and integration of sensory information. Neuroimaging meta-analyses have consistently reported decreased gray matter volume and altered connectivity in the insular cortex in both schizophrenia and bipolar disorder \cite{bipolar-schizo, insula}. The insula is a major node of the salience network, which coordinates switching between cognitive control networks and the default mode network. Importantly, the centrality of the insula within the identified networks supports its role as a transdiagnostic hub, mediating the integration between internally and externally oriented processes, and potentially contributing to both psychotic and affective dysregulation \cite{cognitive}.
    \item Several frontal areas appearing among the highlighted features, including inferior frontal and orbitofrontal regions, are associated with executive control, emotional regulation, and social cognition \cite{friedman2021role}. Structural and functional abnormalities in frontal cortex are among the most consistent findings in both schizophrenia and bipolar disorder, although again the alterations tend to be more pronounced and widespread in schizophrenia \cite{bipolar-schizo, frontal}. The differential engagement of frontal regions is consistent with a hierarchical organization of network alterations, in which schizophrenia is characterized by a more extensive disruption of frontally mediated control systems, whereas bipolar disorder involves a more selective modulation within the same circuitry, as previously observed in structural covariance analyses on the same dataset \cite{mri_data}. Taken together, these findings suggest that the identified regions should not be interpreted as isolated loci, but rather as key nodes within distributed networks, whose coordinated disruption defines the structural organization of psychotic and affective disorders.
\end{enumerate}

The node-level centrality measures provide complementary perspectives on how these brain regions participate in the gray matter correlation network:
\begin{enumerate}[label=(\roman*)]
    \item Strength and eigenvector centrality highlight nodes embedded within dense correlation structures. The consistently elevated values for several regions in the schizophrenia task suggest the presence of strongly coupled correlation hubs, indicating that structural changes in these regions are tightly linked with alterations across multiple cortical areas. In the bipolar task, the same regions remain important, showing comparable centrality values. This is consistent with neuroimaging literature indicating that bipolar disorder shares core structural alterations with schizophrenia but with less extensive network disruption \cite{bipolar-schizo}. This interpretation is further supported by previous analyses on the same dataset \cite{mri_data}, where schizophrenia was associated with multiple altered structural covariance networks, whereas bipolar disorder showed a more circumscribed pattern within overlapping regions.
    \item PageRank centrality reflects how correlations propagate through the network via indirect pathways. The relatively small variability across runs suggests that these regions maintain stable diffusion accessibility, implying that they form part of a persistent communication backbone within the correlation network.
    \item Current-flow betweenness identifies nodes that mediate the flow of correlations between distant regions. The presence of elevated values for certain temporal and frontal regions indicates that they act as topological bridges within the learned interaction graph  between otherwise weakly connected cortical modules, a property that may reflect abnormal integration between sensory and cognitive processing circuits in psychosis \cite{SUZUKI200241,SORELLA2019101854}.
    \item Closeness centrality measures the global accessibility of nodes. The relatively small differences observed across tasks indicate that although local connectivity patterns change, the global geometric position of these nodes within the network remains stable.
\end{enumerate}

A consistent pattern emerges across both quantum-inspired measures analyzed in this work: the single-site entanglement entropy (Appendix~\ref{sec:quantum_inspired_interpretability}) and the network centrality measures (Sec.~\ref{sec:node_level_quantities}) both display substantially higher stability across resampling runs than Shapley values (Appendix~\ref{sec:quantum_inspired_interpretability}). While the stability of the entanglement entropy reflects the structural properties of the global MPS representation, the stability of network centralities has a distinct origin:
centrality measures depend on the second-order statistical structure of the data (pairwise correlations), which is relatively stable under repeated sampling, whereas Shapley values depend on the specific trained model and contextual learned feature interactions. As a result, centrality measures capture structural properties of the feature network, while Shapley values capture model-specific attribution patterns that may fluctuate depending on training conditions. The fact that the same brain regions emerge as stable network hubs despite fluctuations in Shapley scores reinforces the interpretation that these areas represent robust neuroanatomical correlates of the classification tasks.

The node-level analysis supports a model in which the bipolar classification corresponds to a reduced core subnetwork embedded within a broader schizophrenia-related structural network. This hierarchical relationship aligns with neuroimaging studies indicating that schizophrenia typically involves more widespread cortical alterations, whereas bipolar disorder tends to affect a more restricted set of limbic and frontal regions \cite{Dobri2022Limits}. This broader hierarchical perspective also resonates with morphology-based approaches revealing how brain network organization is structured in autism  \cite{Safari2025.08.28.672884}.

Within this framework, the five shared regions identified in the bipolar classification may represent a core structural module common to both disorders, potentially associated with cognitive deficits and sensory integration. The additional regions identified in the schizophrenia task may instead reflect more extensive disruptions of sensory, cognitive, and associative networks, consistent with the broader symptomatology of schizophrenia.

Insular and fronto-temporal regions define a shared structural backbone across both disorders \cite{ELLISONWRIGHT20101}, but with different weights: while the insula reflects a relatively stronger involvement in affective processing typical of bipolar disorder, fronto-temporal and auditory regions exhibit a more pronounced and widespread centrality in schizophrenia, consistent with its broader cortical dysconnectivity profile \cite{schizo-bipolar2015}. These characterizations partially support the asymmetry observed in Fig. \ref{fig:boxplot} about sensitivity and specificity.

The results suggest that graph-theoretic analysis of the learned gray matter correlations networks, obtained from quantum observables, provides a robust framework for identifying stable neurobiological signatures of brain disorders. By focusing on network topology rather than individual feature contributions, the approach reveals hierarchical relationships between disorders and highlights regions that emerge as structural hubs within the pathological network.
In this sense, the integration of graph-based and model-based interpretability may represent a promising direction for developing more robust and interpretable neuroimaging markers.

\section{Conclusion}\label{sec:conclusion}

We introduced a tensor network machine‑learning framework that maps gray‑matter features into quantum‑inspired correlation structures, enabling their interpretation as weighted graphs.
Crucially, the quantum correlations encoded in this framework should not be interpreted as standard measures of statistical covariance. While conventional structural covariance networks rely on pairwise correlations that quantify isolated co-fluctuations, the MPS representation learns dependencies jointly within a single structured model. In this setting, the relationship between any two regions depends on the full configuration of features, rather than being estimated independently.
As a result, although the computed quantum observables take the form of two-point correlations, they are derived from a global model and therefore implicitly capture context-dependent, higher-order interactions. This provides a more expressive and integrated description of feature interactions reflected in the network topology.

Within this unified representation, learning is not only assessed in terms of predictive performance, but also characterized through the emergence of structured correlations, providing direct access to the relational organization of brain features.

By combining MPS with graph‑theoretic analysis, we showed that the learning dynamics are reflected in a sharp topological reorganization of the correlation network. This transition is characterized by global spectral observables, which reveal the formation of coherent collective modes, and by node-level centrality measures, which identify a stable set of brain regions acting as hubs of the learned representation. 

By focusing on network topology rather than isolated feature contributions, we highlighted the emergence of a hierarchical organization of discriminative features, in which the regions relevant for bipolar disorder form a subset of those identified for schizophrenia. This nested structure provides a network-based characterization of the relationship between the two disorders and is consistent with the broader and more distributed structural alterations observed in schizophrenia \cite{mri_data}. 

Importantly, observables encoding quantum correlations, both the entanglement entropy and the graph-based centrality measures, exhibit higher stability across train-test resampling than model-based attribution measures, underscoring the value of quantum-inspired representations for identifying robust neurobiological markers.

Overall, our results demonstrate that tensor‑network models, when coupled with interpretable graph-theoretic analysis, offer a principled framework for uncovering hierarchical and disorder‑specific brain signatures, providing a bridge between quantum‑inspired machine learning and neuroscience. More broadly, these findings support the view that brain disorders may be more effectively characterized through distributed organizational patterns emerging within learned representations than through isolated regional alterations alone.

\begin{acknowledgments}
Authors were supported by the Italian funding within the “Budget MUR - Dipartimenti di Eccellenza 2023 - 2027” (Law 232, 11 December 2016) - Quantum Sensing and Modelling for One-Health (QuaSiModO), CUP:H97G23000100001. G.M. acknowledges support from INFN through the
project “QUANTUM”.

\end{acknowledgments}

\appendix

\section{\label{app:sec1}Training dynamics of the MPS classifier}

In this appendix, we characterize the learning dynamics observed in a representative run of each of the two
binary classification tasks involving clinical data: \emph{schizophrenic vs.\ control subjects} and \emph{bipolar vs.\ control subjects}. For each task, we track the evolution of the evaluation metrics, the reduced density matrix in label space, the magnetization patterns extracted by the two trained MPS weight tensors with bond dimension $\chi=400$, and the entanglement entropy profiles along the one‑dimensional lattice. 

\begin{figure*}
\includegraphics[width=0.95\linewidth]{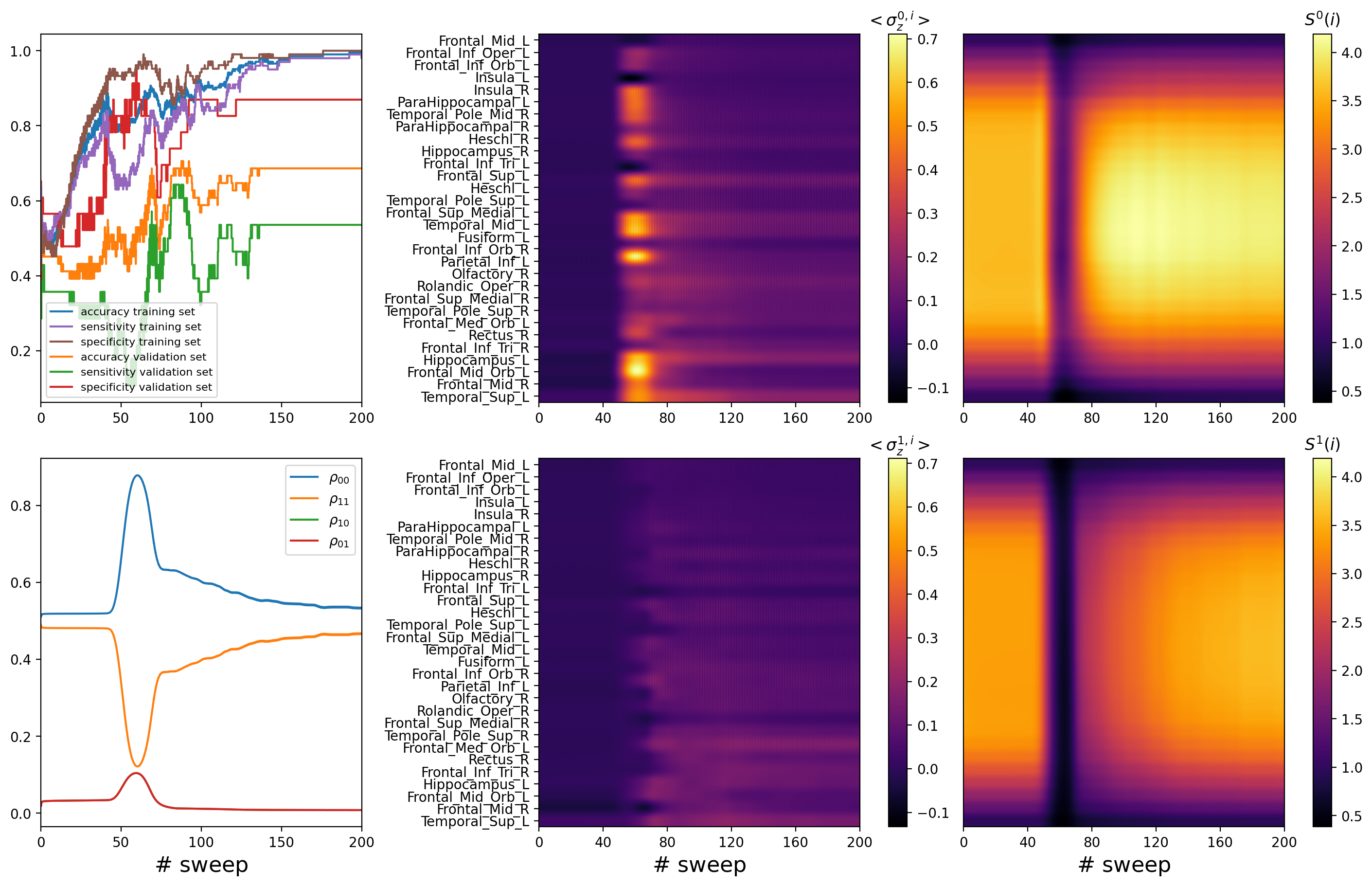}
\caption{\label{fig:app1}
Left panels: evaluation metrics for training and independent sets (top) and reduced density matrix in the label subspace (bottom). 
Central panels: magnetization patterns extracted by the MPS masks for the control class
(top) and schizophrenic patients (bottom). Right panels: entanglement entropy profiles $S^{(0)}(i)$ (controls, top) and $S^{(1)}(i)$ (patients, bottom) along the one-dimensional lattice.}
\end{figure*}

\subsection{Schizophrenic patients vs.\ controls}

In the schizophrenia–control task, shown in Fig.~\ref{fig:app1}, the left column displays the evolution of accuracy, sensitivity, and specificity for both training and independent test sets. The evaluation metrics show a delayed generalization, stabilizing at high values only after a transient regime of fluctuations. Around a relatively early sweep number, which we refer to as the transition sweep, the sensitivity and specificity of the independent test set start to improve and stabilize. This stabilization marks the onset of effective generalization.

This delayed generalization is accompanied by a sharp transition in the internal observables of the trained state, with a phenomenology qualitatively similar to the grokking transition reported in tensor-network classifiers \cite{pomarico2025grokkingentanglementtransitiontensor}. In particular, the reduced density matrix in the label subspace, shown below the metrics, exhibits a sharp reorganization: its off‑diagonal terms decay abruptly around the same sweep, indicating the emergence of two dynamically decoupled masks, and signaling a clear reorganization of the classification subspace.

The central column illustrates the local magnetization patterns for the two trained quantum states: the upper panel corresponds to control subjects, the lower to schizophrenic patients. Features relevant for each class become sharply activated near the transition sweep, forming stable and class‑specific magnetization profiles.

The right column presents the bipartite entanglement entropy $S^{(\ell)}(i)$ across the one‑dimensional lattice for each class. Starting from a volume‑law profile associated with a random initialization, the entropy rapidly collapses into a sub‑volume scaling after the transition sweep. This transition is consistent with the reduced number of effective degrees of freedom required once the model has extracted the minimal feature set necessary for generalization.

The schizophrenia-control classification undergoes this transition at a comparatively \emph{small} number of sweeps, indicating that the model can quickly isolate a concise set of discriminative features.

\subsection{Bipolar patients vs.\ controls}

\begin{figure*}
\includegraphics[width=0.95\linewidth]{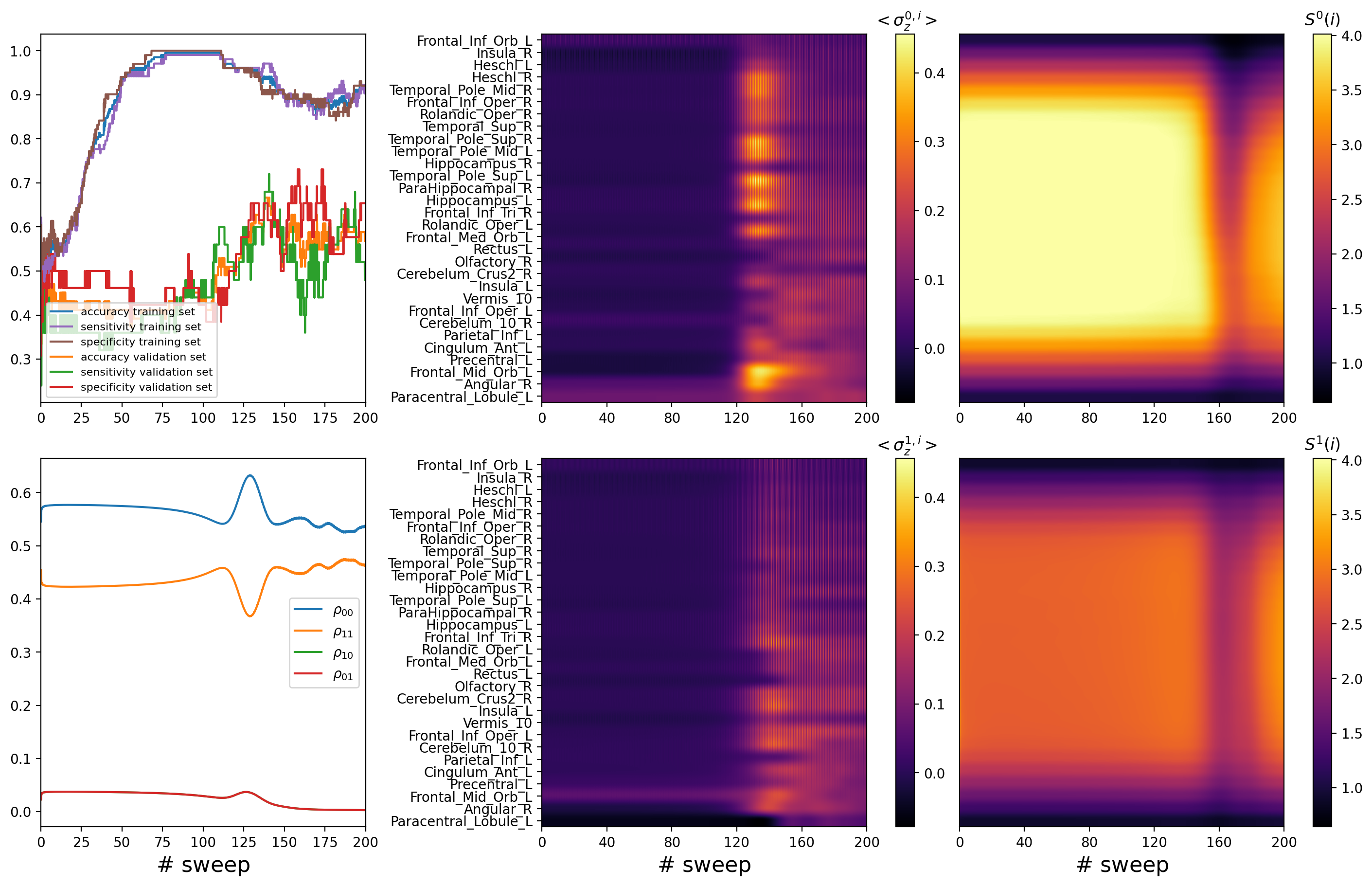}
\caption{\label{fig:app2}
Left panels: evaluation metrics for training and independent sets (top) and reduced density matrix in label space (bottom). Central panels: magnetization patterns for the control class (top) and bipolar patients (bottom). Right panels: entanglement entropy profiles
$S^{(0)}(i)$ (controls, top) and $S^{(1)}(i)$ (patients, bottom).} 
\end{figure*}

The bipolar-control task, shown in Fig.~\ref{fig:app2}, exhibits the same qualitative phenomenology but with a markedly different timescale. In the left column, the accuracy and sensitivity curves for the independent set reveal a delayed generalization phase: the sudden improvement occurs only after a considerably larger number of sweeps compared to the schizophrenia-control case, indicating that the model must explore a more complex landscape before isolating class‑specific features.

The magnetization patterns shown in the central panels corroborate this observation. In contrast to the schizophrenia task, where a small number of regions dominate early, the bipolar-control masks develop discernible patterns only after extended training. This suggests that the discriminative structure of the data is either more diffuse or encoded in more subtle correlations that require more optimization steps to uncover.

The entanglement entropy profiles in the right column again begin in a volume‑law regime. However, the collapse to a sub‑volume profile occurs later, precisely aligned with the delayed transition point. The fact that the volume-to-sub-volume transition is postponed reflects the need for the MPS to retain higher entanglement for a longer period while navigating a larger or more entangled feature manifold before locking onto the relevant discriminatory subspace.

\subsection{Comparative interpretation}

The two sets of panels illustrate that although both tasks display the characteristic signatures of delayed generalization, i.e. decay of label space coherence, emergence of stable magnetization, and an entanglement entropy transition, the \emph{timescales} of these transitions differ markedly.

The schizophrenia vs. control task reaches high generalization after relatively few sweeps, indicating that its feature structure is easier to isolate within the MPS architecture.  
In contrast, the bipolar–control task requires significantly more sweeps, implying that its discriminatory features are harder to extract, more entangled, or less spatially localized along the one-dimensional lattice.

\begin{figure}[t]
\includegraphics[width=0.95\linewidth]{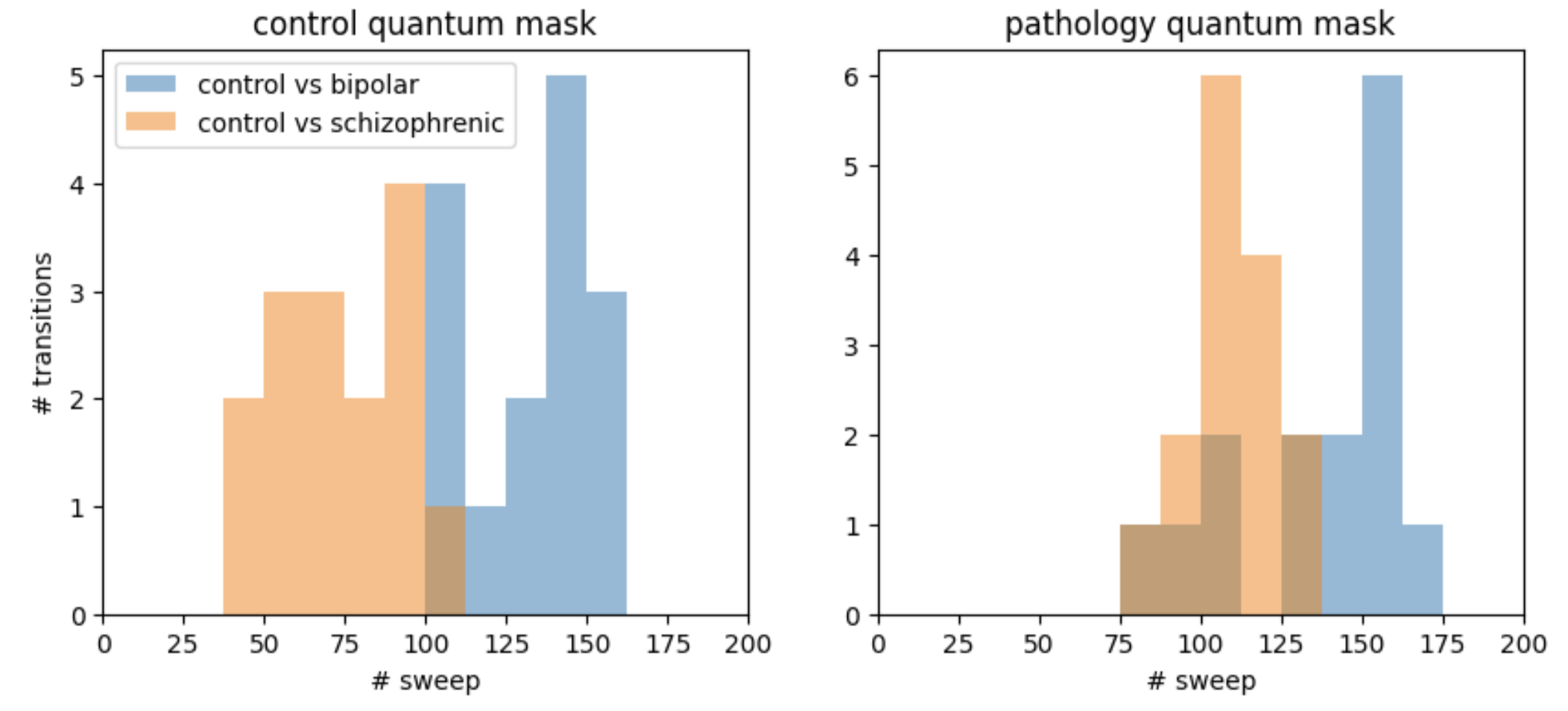}\caption{\label{fig:grokking_hist} Distribution of transition sweeps collected over 15 independent training–test resampling runs for the two diagnostic tasks. Left: transition–sweep histograms for the control quantum state in the bipolar–control (blue) and schizophrenia–control (orange) tasks.  Right: corresponding histograms for the pathology quantum states.  
The schizophrenia–control task consistently exhibits earlier and more tightly clustered transitions, while the bipolar–control task shows later and more variable transition times, reflecting the differing complexity of the underlying feature structures.}
\end{figure}

To assess the stability of these observations against dataset variability, we performed 15 independent training–test resampling runs for each diagnostic task. The resulting histograms of transition sweeps, shown in Fig.~\ref{fig:grokking_hist}, provide a quantitative measure of the variability of the transition times for both the control and pathology quantum states.

For the schizophrenia–control comparison, the distribution of transition sweeps (orange bars) is consistently shifted toward earlier values. In the control histogram, the schizophrenia transitions cluster sharply in the range $\sim 60$-$110$ sweeps, with a narrow spread and no heavy tail. The analogous histogram for the state encoding the pathology class exhibits the same early‑transition behaviour, indicating that the onset of stable magnetization and entanglement reorganization is reproducible across resampling and not driven by a specific partition of the data.

In contrast, the bipolar–control task (blue bars) exhibits systematically later transitions and a markedly broader spread. Both the control and pathology states show transitions concentrated primarily between $\sim 120$-$170$ sweeps, with higher variance and a noticeable right‑skew.  
This broader distribution signals that the transition dynamics are more sensitive to resampling, consistent with the interpretation that bipolar features are intrinsically harder to disentangle and require more extensive training for the masks to isolate class-specific signatures.

The histogram analysis reinforces the comparative picture: schizophrenia–control classifications consistently undergo transitions early and with low variance across resampling, whereas bipolar–control classifications transition substantially later, with higher trial‑to‑trial variability. This behaviour mirrors the differences already evident at the level of magnetization patterns, reduced density matrix decoherence, and entanglement‑entropy transitions, confirming that the observed timescale separation is a robust phenomenon rather than a single‑run artefact.

\section{\label{sec:quantum_inspired_interpretability}Quantum-inspired interpretability}

\begin{figure*}[t]
    \centering
    \includegraphics[width=0.9\linewidth]{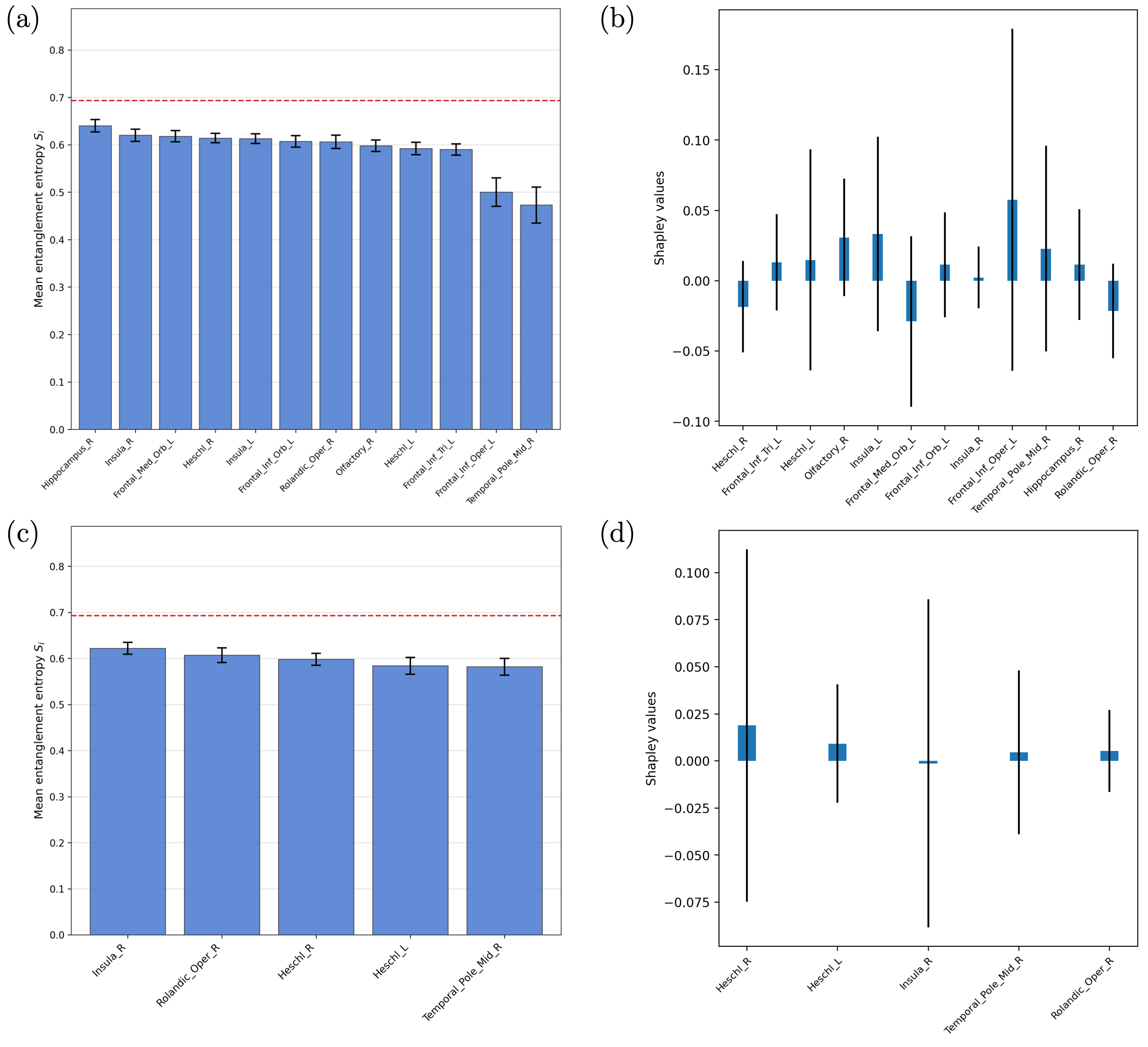}
    \caption{ Quantum entropy and standard interpretability measures for the control vs. schizophrenia classification (panels a–b) and control vs. bipolar classification (panels c–d). Panels (a) and (c) show the mean single-site entanglement entropy $S_i$ computed from the trained MPS, averaged over the 15 resampling runs, for the features consistently selected by Boruta across all runs. The red dashed line indicates the maximum achievable value of the single-site entanglement entropy. Panels (b) and (d) show the Shapley values assigned to the same subset of regions, reflecting each region's contribution to the XGBoost classifier predictive decision.}
    \label{fig:entropy_feature_importance}
\end{figure*}

In order to extract feature information from the trained quantum-inspired model, we consider the single-site entanglement entropy \cite{Felser_2021, AIZPURUA2025130211}. For each site $i$ of the MPS, we compute the reduced density matrix $\varrho_i$, obtained by tracing out all other sites from the trained MPS and summing over the class-label index so that $\varrho_i$ encodes information from both classes simultaneously. The single-site entanglement entropy is then defined as $S_i = -\textrm{Tr}\left ( \varrho_i \textrm{ln} \varrho_i \right )$, which quantifies the degree of quantum correlation between the feature $i$ and  the other features within the MPS representation. Since each feature is encoded as a qubit, $S_i$ is bounded above by $\textrm{ln}2  \approx 0.693$. Features with higher entanglement entropy are more deeply integrated into the collective quantum correlations that define the learned representation. Importantly, the entropy values should be interpreted as properties of the learned representation rather than direct neurobiological measures. By construction, this measure reflects representational integration rather than marginal contribution to the classification output, which is the focus of attribution methods such as Shapley values \cite{shapley1951notes, lundberg2017unified}. We compute the single-site entanglement entropy from the trained MPS and the Shapley values from the XGBoost baseline, obtaining both quantities for the same set of features and treating them as complementary probes of the learned structures. 

In Fig.~\ref{fig:entropy_feature_importance}, we report the mean single-site entanglement entropy $S_i$ (panels a and c), and the Shapley values (panel b and d) for the features consistently selected by Boruta across all 15 resampling runs, i.e. 12 stable regions for the schizophrenia task and 5 for the bipolar task. Both quantities are averaged over the 15 runs. 

A key observation emerges from the comparison with the Shapley values, which show greater variability between runs for both classification tasks. This indicates that the single-site entanglement entropy is more stable across resampling runs than the model-based attribution scores. While Shapley values depend on the specific trained model and learned decision boundary, the entanglement entropy remains robust under data splits.

\section{\label{app:sec2}Graph theory global and node-level quantities}

We introduce in this appendix the weighted graph quantities employed in the analysis of the structural organization underlying the quantum masks $W^\ell$ adopted for the binary classification tasks.

Let $G = (V,E)$ be a weighted undirected graph with $|V| = N$ nodes and 
weighted adjacency matrix $A = (a_{i,j}) \in \mathbb{R}^{N\times N}$, where 
$a_{i,j} =|C_{i,j}| \ge 0$ encodes the strength of the interaction between nodes $i$ and $j$, where $C_{i,j}$ is introduced in Eq. \eqref{eq:connectedcorr}.

\subsection{Global quantities}

\subsubsection{Largest eigenvalue of the weighted adjacency matrix}

The spectral radius of $A$,
\begin{equation}
\lambda_N(A) = \max \{\, \mathrm{spec}(A) \,\},
\end{equation}
provides a global descriptor of the graph's connectivity strength.   For non-negative symmetric matrices, $\lambda_N(A)$ coincides with the  largest real eigenvalue and is associated with a strictly non-negative eigenvector by the Perron-Frobenius theorem. In weighted graphs this quantity captures the presence of strongly connected hubs or dense subgraphs and is often interpreted as a proxy for the network’s overall
interaction intensity \cite{Chung1997}.

\subsubsection{Spectral entropy of the adjacency matrix}

To quantify the heterogeneity of the adjacency spectrum, we define the normalized squared eigenvalues
\begin{equation}
p_i = \frac{\lambda_i^2}{\sum_{j=1}^N \lambda_j^2},
\qquad i = 1,\ldots,N,
\end{equation}
where $\{\lambda_i\}$ are the eigenvalues of $A$.  
The spectral entropy is then given by \cite{Thomas_2025}
\begin{equation}
H_{\mathrm{spec}}(A)
=
- \sum_{i=1}^N p_i \log p_i .
\end{equation}
A low spectral entropy indicates that a small number of eigenmodes dominate the
connectivity structure (e.g., the presence of few influential hubs), whereas
higher entropy reflects a more uniform distribution of topological scales.

\subsubsection{Algebraic connectivity}

Let $D = \mathrm{diag}(d_1,\ldots,d_N)$ denote the degree matrix with 
$d_i = \sum_j a_{ij}$, and define the weighted Laplacian
\begin{equation} \label{eq:laplacian}
L = D - A .
\end{equation}
Since $L$ is symmetric and positive semidefinite, its eigenvalues satisfy
\begin{equation}
0 = \mu_1 \le \mu_2 \le \cdots \le \mu_N .
\end{equation}
The second smallest eigenvalue, $\mu_2$, is called the \emph{algebraic connectivity}
(or Fiedler value) \cite{Fiedler1973, DEABREU200753}:
\begin{equation}
\alpha(G) = \mu_2 .
\end{equation}
This quantity measures the robustness of the network’s global connectivity: 
larger values indicate that the graph is more difficult to partition and possesses 
more redundant or distributed pathways across its nodes.

\subsubsection{Laplacian spectral ratio}

To capture both the connectivity strength and the spectral spread of the Laplacian,
we define the Laplacian ratio as \cite{PhysRevLett.89.054101, PhysRevE.77.031102, YOU20121245}
\begin{equation}
R_L(G) = \frac{\mu_N}{\mu_2},
\end{equation}
where $\mu_N$ is the largest eigenvalue of $L$.  
A high Laplacian ratio reflects a network with tightly clustered or highly 
heterogeneous connectivity patterns, while smaller values indicate a more 
homogeneous connectivity profile.  
This ratio is particularly informative when comparing networks that share similar
densities but differ in structural organization, as it incorporates both global
cohesion ($\mu_2$) and maximal variation across nodes ($\mu_N$).

\subsection{Node-level quantities}

\subsubsection{Node strength}

For a weighted, undirected network with adjacency matrix $A=(a_{ij})$, the
\emph{strength} of node $i$ is the sum of incident edge weights \cite{pnas.0400087101}
\begin{equation}
s_i \;=\; \sum_{j=1}^N a_{i,j}.
\end{equation}
The vector $\bm{s}=(s_1,\ldots,s_N)^\intercal$ equals $A\mathbf{1}$, where $\mathbf{1}$
is the all-ones vector. Node strength generalizes degree to weighted graphs and quantifies the total interaction intensity of a node in terms of its connectivity.

\subsubsection{Eigenvector centrality}

Eigenvector centrality assigns high score to nodes that are connected to other
highly central nodes, according to each component value $x_i, \, i=1, \dots ,N$. It is defined as the (entrywise nonnegative) eigenvector associated with the largest eigenvalue of $A$ \cite{Newman2018}
\begin{equation}
A\, \bm{x} \;=\; \lambda_N (A)\, \bm{x}, \qquad x_i \ge 0,\ \ \| \bm{x} \|_2=1.
\end{equation}
In weighted graphs, edges with larger $a_{ij}$ contribute more strongly to the mutual reinforcement among neighboring centralities.

\subsubsection{PageRank}

Let $P$ be a row-stochastic transition matrix derived from $A$. A possible choice
for undirected weighted graphs is $P_{i,j} \;=\; 
a_{i,j}/s_{i}$, which performs a random walk proportional to edge weights. We can introduce a centrality satisfying a steady state condition $\bm{\pi}^\intercal \;=\; \bm{\pi}^\intercal P$, yielding the normalized strength $\pi_i \;=\; s_i/\sum_k s_k$.

The PageRank vector \cite{Page1999} is deduced by introducing a uniform fallback because dense weighted graphs are often nearly fully connected and easily present dominant eigenmodes. A damping factor $\alpha \in (0,1)$ weights the inclusion of a personalization vector $\bm{v}$ (assumed uniform in our analysis, $\bm{v}=\tfrac{1}{N}\mathbf{1}$), that biases the random surfer to restart at specific, preferred nodes, thus customizing importance scores, essential for personalized recommendations, user-specific search and community detection. The steady state satisfies
\begin{equation}
\bm{\pi}^\intercal = \alpha \bm{\pi}^\intercal P + (1-\alpha) \bm{v}^\intercal,
\end{equation}
where larger $\alpha$ emphasizes network structure, while smaller
$\alpha$ increases the influence of $\bm{v}$. In our implementation the damping factor $\alpha \;=\; 0.85$, ruling the transition matrix $\widetilde{P} = \alpha P + \dfrac{1-\alpha}{N} \mathbf{1} \mathbf{1}^\intercal$.

\subsubsection{\label{app:subsecH}Current-flow betweenness}

Current-flow betweenness \cite{978-3-540-31856-9_44, BOZZO2013460} measures how often a node carries flow when one unit of current is injected at a source node $s$ and extracted at a target one $t$, assuming each edge $(i,j)$ has conductance $a_{i,j}$ (thus resistance
$r_{i,j}=1/a_{i,j}$ for $a_{i,j}>0$). Let $L$ be the weighted Laplacian introduced in Eq. \eqref{eq:laplacian} and $L^+$ its Moore-Penrose pseudoinverse. 

Ohm’s law on edge $(i,j)$ reads $I_{i,j}^{s,t} = a_{i,j} (v_i^{s,t} - v_j^{s,t})$, where $v_i^{s,t}$ is the voltage at node $i$ and $I_{i,j}^{s,t}$ the current from $i$ to $j$. Kirchhoff’s current law at each node $i$ yields the external current injection $\sum_{j} I_{i,j}^{s,t}=b_i^{s,t}$, and the application of Ohm’s law results in
\begin{equation}
    v_i^{s,t} \sum_{j} a_{i,j} - \sum_{j} a_{i,j} v_j^{s,t} = \sum_j L_{i,j} v_j^{s,t} = b_i^{s,t}.
\end{equation}

For a given pair $(s,t)$ with $b_i^{s,t}=e^s_i - e^t_i$, the node potential vector is $v_i^{s,t}= L^+ (e_i^s - e_i^t)$, where $e_i$ is the
$i$-th canonical basis vector. The net current through node $i$ induced by the unit $s\!\to\! t$ injection can be expressed in terms of potential differences across incident edges. The current-flow betweenness of node $i$ is then the average of its absolute through-current over all ordered pairs:
\begin{widetext}
\begin{equation}
\mathrm{B}(i)
\;=\;
\frac{1}{N(N-1)}
\sum_{\substack{s,t \in V\\ s\neq t}}
\frac{1}{2}\!
\sum_{j}
a_{ij}\,\bigl| \, L^+ (e_i^s - e_i^t)
- L^+ (e_j^s - e_j^t) \,\bigr|,
\end{equation}
\end{widetext}
where the inner sum aggregates the magnitude of current traversing node $i$
via edges $(i,j)$. In practice, stochastic sampling over $(s,t)$ pairs can be used to estimate $\mathrm{B}$ efficiently in large networks.

\subsubsection{Closeness}

We define a geometric distance between nodes from the edge weights as
\begin{equation}
d_{i,j} \;=\; \sqrt{\, 2(1 - a_{i,j})\,}\,,
\qquad d_{i,i}=0,
\label{eq:dist_def}
\end{equation}
so that larger weights imply shorter distances. This Mantegna distance formula comes from embedding standardized variables on a unit hypersphere. The
weighted closeness centrality of node $i$ is defined as \cite{10.1121/1.1906679, Sabidussi_1966}
\begin{equation}
c_i \;=\; \frac{N-1}{\displaystyle \sum_{j\neq i} d_{i,j}}.
\label{eq:weighted_closeness}
\end{equation}
where for pairs without a direct edge, we set $d_{i,j}$ to the length of the \emph{shortest path} under Eq. \eqref{eq:dist_def},
i.e.\ the minimal path sum of edge-wise distances along any $i{\to}j$ route
in the graph (finite if $j$ is reachable from $i$, and $+\infty$ otherwise).

\section{\label{app:sec3} Supplementary figures}

\begin{figure*}[t]
\subfigure[]{\includegraphics[width=0.35\linewidth]{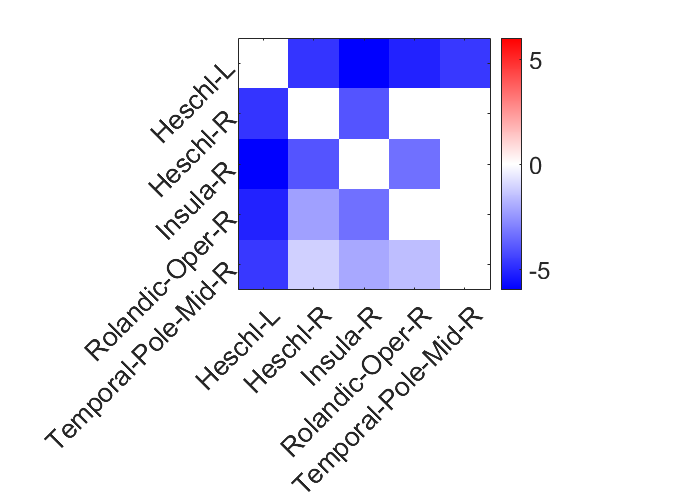}}
\subfigure[]{\includegraphics[width=0.5\linewidth]{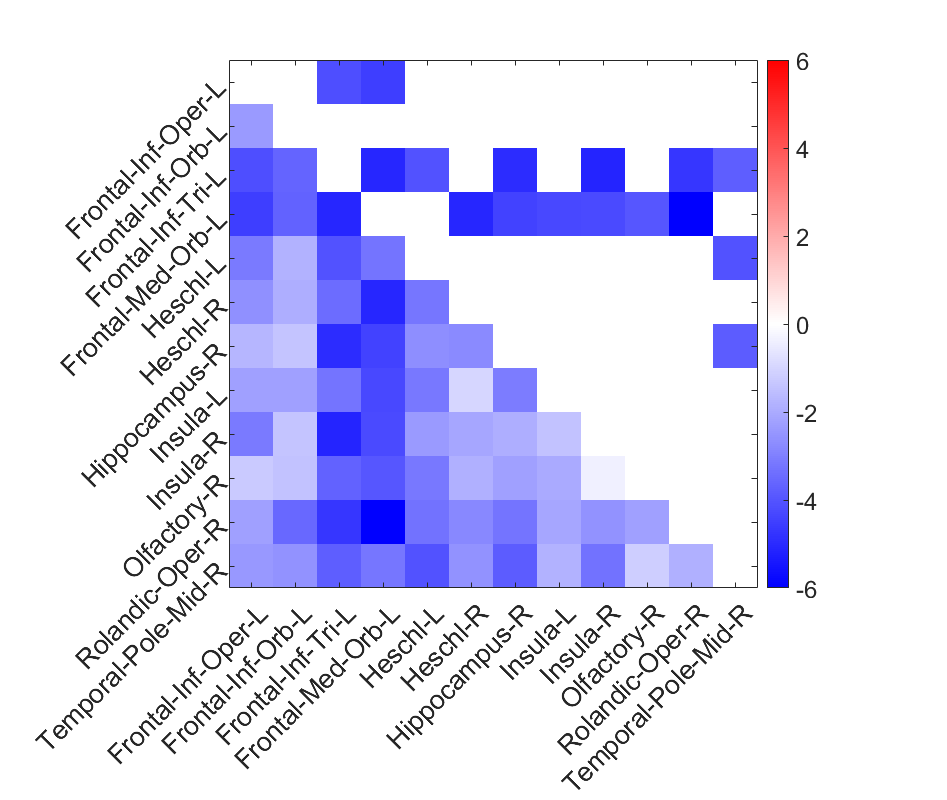}}
\caption{\label{fig:bip_sz_mt} T-values for group differences in correlation networks: controls versus bipolar disorder (a) and schizophrenia (b). The upper triangle of each matrix indicates the pairwise regions with statistically significant differences.}
\end{figure*}

\bibliography{apssamp}% Produces the bibliography via BibTeX.

\end{document}